\documentclass[11pt]{article} 
\usepackage{a4}
\usepackage{a4wide}
\usepackage{amssymb}
\usepackage{amsmath}
\usepackage{color}
\usepackage{array}
\usepackage{graphics,epsfig}
\usepackage{graphicx}
\usepackage{amsfonts}
\usepackage{psfrag}
\newcommand{\munite}{\ensuremath{\,\,\mathrm{l}\!\!\!1}}
\newcommand{\ov}{\overline}

\newtheorem{prop}{Property}

\title{From modular invariants to graphs: the modular splitting method}

\author{{\bf E. Isasi}${}^{1,4}$\thanks{Email: eisasi@usb.ve} , {\bf G. Schieber}${}^{2,3}$\thanks{Supported by a fellowship of AUF - Agence Universitaire de la Francophonie, and of FAPERJ, Brasil. E-mail: schieber@cbpf.br} .  \\
\\
  ${}^1$ {\it CPT - Centre de Physique Th\'eorique\thanks{Unite Mixte de Recherche (UMR 6207) du CNRS
et des Universit\'es Aix-Marseille I, Aix-Marseille II,
et du Sud Toulon-Var; laboratoire affili\'e \`a la
FRUNAM (FR 2291).}} \\
  {\it Campus de Luminy - Case 907}\\
  {\it F-13288 Marseille - France}\\
\\
${}^2$ {\it Laboratoire de Physique Th\'eorique et des Particules (LPTP)} \\ 
{\it D\'epartement de Physique, Facult\'e des Sciences} \\
                 {\it Universit\'e Mohamed I, B.P.524 Oujda 60000, Maroc}           \\
\\
${}^3$ {\it CBPF - Centro Brasileiro de Pesquisas F\'{\i}sicas} \\
                 {\it Rua Dr. Xavier Sigaud, 150}\\
                 {\it 22290-180, Rio de Janeiro, Brasil} \\
\\
${}^4$ {\it Departamento de F\'{\i}sica}\\
       {\it Universidad Sim\'on Bol\'{\i}var}\\
       {\it AP 89000, Caracas 1080-A, Venezuela}
}

\date{}
\begin{document}
\thispagestyle{empty}

\begin{titlepage}

\maketitle


\begin{abstract}
We start with a given modular invariant $\mathcal{M}$ of a two dimensional $\widehat{su}(n)_k$ conformal field theory (CFT) and present a general method for solving the Ocneanu modular splitting equation  and then determine, in a step-by-step explicit construction, 1) the generalized partition functions corresponding to the introduction of boundary conditions and defect lines; 2) the quantum symmetries of the higher ADE graph $G$ associated to the initial modular invariant $\mathcal{M}$. Notice that one does not suppose here that the graph $G$ is already known, since it appears as a by-product of the calculations.  
We analyze several $\widehat{su}(3)_k$ exceptional cases at levels 5 and 9. 
\vspace{1.0cm}
\end{abstract}

\vfill

\noindent {\bf Keywords}: conformal field theory, modular invariance, higher ADE systems, fusion algebra, Hopf algebra, quantum groupo\"ids.



\vspace*{0.5 cm}

\end{titlepage}



\section{Introduction}
Following the works of \cite{Oc-paths}, it was shown that to every modular invariant of a 2d CFT one can associate a special kind of quantum groupo\"{\i}d $\mathcal{B}(G)$, constructed from the combinatorial and modular data \cite{ganon-moddata} of a graph $G$ \cite{Pet_Zub-Occells,Coq_Trinchero,Gil-tesis,Trinchero,Coq-q6j}. This quantum groupo\"{\i}d $\mathcal{B}(G)$ plays a central role in the classification of 2d CFT, since it also encodes information on the theory when considered in various environments (not only on the bulk but also with boundary conditions and defect lines):  the corresponding generalized partition functions are expressed in terms of a set of non-negative integer coefficients that can be determined from associative properties of structural maps of $\mathcal{B}(G)$ \cite{behrend-BCFT,Zub-Bariloche,Pet_Zub-gener,Gil-tesis}. A series of papers \cite{Pet_Zub-gener,Coq-qtetra, Pet_Zub-Occells, Coq_Gil-ADE, Coq_Gil-Tmod,Gil-tesis} presents the computations allowing to obtain these coefficients from a general study of the graph $G$ and its quantum symmetries. In this approach, the set of graphs $G$ is taken as an input. For the $\widehat{su}(2)_k$ model, the graphs $G$ are the ADE Dynkin diagram, and  for the $\widehat{su}(3)_k$ the Di Francesco-Zuber diagrams. A list of graphs has also been proposed in \cite{Oc-Bariloche} for the $\widehat{su}(4)_k$ model. For a general $SU(N)$ system, the set of graphs $G$ presents the following pattern. There is always the infinite series of $\mathcal{A}_k$ graphs, which are the truncated Weyl alcoves at some level $k$ of $SU(N)$ irreps.  Other infinite series are obtained by orbifolding and conjugation methods, but there are also some exceptional graphs (generalizing the $E_6$ and $E_8$ diagrams of the $SU(2)$ series) that can not be obtained in that way (to some extent, the $E_7$ diagram can be obtained from a careful study of the $D_{10}$ case). One of the purposes of this article is actually to present a method to obtain these graphs. \\

We start with a modular invariant of a 2d $\widehat{su}(n)_k$ CFT as initial data. Classification of modular invariants is only completed for $n=2$ and 3, but there exist several algorithms, mostly due to T. Gannon, that allow one to obtain modular invariants up to rather high levels of any affine algebra. By solving the modular splitting equation (to be recalled later), we obtain the coefficients of the generalized partition functions, as well as the quantum symmetries of the graph $G$, encoded in the Ocneanu graph $Oc(G)$. The graph $G$ itself is then obtained at this stage as a subgraph or a module graph of its own Ocneanu graph: it appears as a by-product of the computations.

Notice that the determination of the higher ADE graphs $G$ by solving the modular splitting equation seems to be the method followed by A. Ocneanu (see \cite{Oc-msri}) to obtain the lists of $SU(3)$ and $SU(4)$ graphs presented in \cite{Oc-Bariloche}, 
as a generalization of the Xu's algorithm \cite{Xu} (see also \cite{Pet_Zub-bound}). 
But explicitation of his method was never been made available in the literature. The method that we describe here (that incorporates the solution of the modular splitting equation itself) was briefly presented in \cite{Coq_esteban} for the study of the non simply laced diagram $F_4$, and is extended and presented in more general grounds.\\

The paper is organized as follows.  In section {\bf 2} we review some results of CFT in order to fix our notations, and present the basic steps of the method allowing to solve the modular splitting equation. Section {\bf 3} treats with more technical details of the resolution, making the difference between commutativity or non commutativity of the quantum symmetry algebra. In the last section we analyze some examples in order to illustrate the techniques. First we treat two exceptional $SU(3)$ modular invariants at level 5, labelled by the graphs $\mathcal{E}_5$ and $\mathcal{E}_5/3$. 
The last example is the level 9 exceptional $SU(3)$ modular invariant, which is a special case since it leads to a non-commutative algebra of quantum symmetries and that there are two 
different graphs, $\mathcal{E}_9$ and $\mathcal{E}_9/3$, associated to it. We also discuss the third graph initially associated to the same modular invariant in \cite{DiFZuber} but later rejected by Ocneanu in \cite{Oc-Bariloche}.


\section{CFT and graphs}

Consider a 2d CFT defined on a torus, where the chiral algebra is an affine algebra $\widehat{su}(n)_k$ at level $k$. The modular invariant partition function reads
\begin{equation} 
\mathcal{Z} = \sum_{\lambda,\mu} \chi_{\lambda} \, \mathcal{M}_{\lambda\mu} \, \overline{\chi_{\mu}} \;,
\end{equation} 
where $\chi_{\lambda}$ is the character of the element $\lambda$ of the finite set  of integrable representations of $\widehat{su}(n)_k$, and where the matrix $\mathcal{M}$ is called the modular invariant: it commutes with the generators $\mathcal{S}$ and $\mathcal{T}$ of the modular group $PSL(2,\mathbb{Z})$. 
The introduction of boundary conditions (labelled by $a,b$), defect lines (labelled by $x,y$) or the combination of both, result in the following generalized partition functions (see \cite{cardy-eq,behrend-BCFT,Pet_Zub-gener}):
\begin{eqnarray}
\label{part1} \mathcal{Z}_{a|b}  &=& \sum_{\lambda} (\mathcal{F}_{\lambda})_{ab} \chi_{\lambda} \\
\label{part2} \mathcal{Z}_{x|y}  &=& \sum_{\lambda,\mu} (\mathcal{V}_{\lambda\mu})_{xy} \chi_{\lambda} \overline{\chi_{\mu}} \\
\label{part3} \mathcal{Z}_{x|ab}  &=& \sum_{\lambda} (\mathcal{F}_{\lambda}\, S_x)_{ab}\, \chi_{\lambda}
\end{eqnarray}
All coefficients appearing in the above expressions express multiplicities of irreducible representations in the Hilbert space of the corresponding theory and are therefore non-negative integers. They are conveniently encoded in a set of matrices: the annular matrices $F_{\lambda}$ with coefficients $(\mathcal{F}_{\lambda})_{ab}$; the double annular matrices $V_{\lambda\mu}$ with coefficients $(\mathcal{V}_{\lambda\mu})_{xy}$ and the dual annular matrices $S_x$ with coefficients $(S_x)_{ab}$. The different set of indices run as $\lambda,\mu = 0,\ldots,d_I-1$; $a,b=0,\ldots,d_G-1$ and $x,y=0,\ldots,d_O-1$. The integer $d_I$ is the number of irreps at the given level $k$; the integers $d_G$ and $d_O$ are given in terms of the modular invariant $\mathcal{M}$ by $d_G=Tr(\mathcal{M})$ and  $d_O=Tr(\mathcal{M}\mathcal{M}^{\dag})$ 
(see \cite{Ost,BEK,evans-pinto}). 

Compatibilities conditions -- in the same spirit than those defined by Cardy in \cite{cardy-eq} for boundary conditions -- impose relations on the above coefficients (see \cite{DiFZuber,behrend-BCFT,Pet_Zub-gener}). Altogether they read: 
\begin{eqnarray}
\label{fusF} F_{\lambda} \, F_{\lambda'} &=& \sum_{\lambda''} \mathcal{N}_{\lambda \lambda'}^{\lambda''} \,  F_{\lambda''} \\
\label{dfe} V_{\lambda\mu} \, V_{\lambda'\mu'} &=& \sum_{\lambda''\mu''} \mathcal{N}_{\lambda \lambda'}^{\lambda''} \, \mathcal{N}_{\mu \mu'}^{\mu''} \, V_{\lambda''\mu''} \\
\label{dualS} S_x \, S_y &=& \sum_{z} \mathcal{O}_{yx}^{z} \,  S_{z}
\label{DFE1}
\end{eqnarray}
$\mathcal{N}_{\lambda\mu}^{\nu}$ are the fusion coefficients describing the tensor product decomposition $\lambda \star \mu = \sum_{\nu} \mathcal{N}_{\lambda\mu}^{\nu} \, \nu$ of representations $\lambda$ and $\mu$ of $\widehat{su}(n)_k$. They can be encoded in matrices $N_{\lambda}$ called fusion matrices. $\mathcal{O}_{xy}^z$ are the quantum symmetry coefficients and can be encoded in matrices $O_x$ called quantum symmetry matrices. 

The matrices $\{F_{\lambda},N_{\lambda},O_x,V_{\lambda\mu},S_x\}$ have non negative integer coefficients: they can be seen as the adjacency matrices of a set of graphs. Knowledge of these graphs helps therefore to the complete determination of the partition functions (\ref{part1}), (\ref{part2}) and (\ref{part3}). All these coefficients also define (or can be obtained by) structural maps of a special kind of quantum groupo\"{\i}d \cite{Oc-paths,Pet_Zub-Occells,Gil-tesis,Coq_Trinchero,Coq-q6j}. It is not the purpose of this paper to explore those correspondences, nor to study the mathematical aspects of this quantum groupo\"{\i}d. What we will do here is to determine, taking as initial data the knowledge of the modular invariant $\mathcal{M}$, all the coefficients of the above matrices.

\subsection{Steps of the resolution}
We start with the double fusion equations (\ref{dfe}), which are matrix equations involving the double annular matrices $V_{\lambda\mu}$, of size $d_O \times d_O$, with coefficients $(V_{\lambda\mu})_{xy}$. Notice that these coefficients can also be encoded in matrices $W_{xy}$, of size $d_I \times d_I$, with coefficients $(W_{xy})_{\lambda\mu} = (V_{\lambda\mu})_{xy}$. The $W_{xy}$ are called double toric matrices. When no defect lines are present ($x=y=0$), we must recover the modular invariant of the theory, therefore $W_{00} = \mathcal{M}$. 
Using the double toric matrices $W_{xy}$, the set of equations (\ref{dfe}) read:
\begin{equation}
\sum_z (W_{xz})_{\lambda\mu} \, W_{zy} = N_{\lambda}\,W_{xy}\,N_{\mu}^{tr} \;.
\label{DFE2}
\end{equation}
The successive steps of resolution are the following:

\paragraph{Step 1: toric matrices}
Setting $x=y=0$ in (\ref{DFE2}) and using the fact that $W_{00}=\mathcal{M}$ we get:
\begin{equation}
\sum_z (W_{0z})_{\lambda\mu} \, W_{z0} = N_{\lambda}\,\mathcal{M}\,N_{\mu}^{tr} \;.
\label{MSE}
\end{equation}
This equation was first presented by A. Ocneanu in  \cite{Oc-Bariloche} and is called the {\bf modular splitting equation}. The r.h.s. of (\ref{MSE}) involves only known quantities, namely the modular invariant $\mathcal{M}$ and the fusion matrices $N_{\lambda}$. The l.h.s. involves the set of toric matrices $W_{z0}$ and $W_{0z}$, that we determine from this equation. 

\paragraph{Step 2: double fusion matrices}
Setting $y=0$ in (\ref{DFE2}) we get: 
\begin{equation}
\sum_z (W_{xz})_{\lambda\mu} \, W_{z0} = N_{\lambda}\,W_{x0}\,N_{\mu}^{tr}
\label{MSEgen}
\end{equation}
Once the toric matrices $W_{x0}$ have been determined from Step 1, the r.h.s. of (\ref{MSEgen}) then involves only known quantities. Resolution of these equations determine the double toric matrices $W_{xy}$ -- and equivalently the double fusion matrices $V_{\lambda\mu}$ -- appearing in the l.h.s. of (\ref{MSEgen}).

\paragraph{Step 3: Ocneanu graph}
The double fusion matrices $V_{\lambda\mu}$ are generated by a subset of fundamental matrices $V_{f0}$ and $V_{0f}$, where $f$ stands for the generators of the fusion algebra (for $SU(n)$ there are $n-1$ fundamental generators). These matrices are the adjacency matrices of a graph called the Ocneanu graph. Its graph algebra is the quantum symmetry algebra, encoded in the set of matrices $O_x$.

\paragraph{Step 4: higher ADE graph $G$}
The higher ADE graph $G$ corresponding to the initial modular invariant $\mathcal{M}$ is recovered at this stage as a module graph of the Ocneanu graph. It may be a subgraph of $Oc(G)$ or an orbifold of one of its subgraphs. One also distinguishes type I cases (also called subgroup or self-fusion cases) and type II cases (also called module or non self-fusion cases).

\paragraph{Step 5: realization of the Ocneanu algebra}
Once the higher ADE graph $G$ has been obtained, and following the works of \cite{Coq-qtetra,Coq_Gil-ADE,Gil-tesis}, we propose a realization of its quantum symmetry algebra $Oc(G)$ as a particular tensor product of graph algebras. Each case being singular, we refer to the examples treated in the last section for more details. This realization allows a simple expression for the matrices $O_x$ and $S_x$.

\paragraph{Comments}
The first three steps of the method presented here can be seen as a generalization of an algorithm proposed by Xu \cite{Xu} for the determination of generalized Dynkin diagrams (see also \cite{Pet_Zub-bound, BEK}). The role of the annular matrix element $(F_{\lambda})_{00}$ in Xu's construction is played here by the partition function multiplicity $\mathcal{M}_{\lambda\mu} = (V_{\lambda\mu})_{00}$. The method described here is more general, allowing the determination of the set of matrices $\{F_{\lambda},N_{\lambda},O_x,V_{\lambda\mu},S_x\}$ and the corresponding graphs.


\section{From the modular invariant to graphs}
We start with a modular invariant $\mathcal{M}$ at a given level $k$ of a $\widehat{su}(n)$ CFT, and the corresponding fusion matrices $N_{\lambda}$. 
\subsection{Determination of toric matrices $W_{x0}$}
We compute the set of matrices $K_{\lambda\mu}$ defined by:
\begin{equation}
K_{\lambda\mu} = N_{\lambda}\,\mathcal{M}\,N_{\mu}^{tr} \;.
\label{klm}
\end{equation}
The modular splitting equation (\ref{MSE}) then reads:  
\begin{equation}
K_{\lambda\mu} = \sum_{z=0}^{d_O-1} (W_{0z})_{\lambda\mu} \, W_{z0} \;.
\label{Kmse}
\end{equation}
This equation can be viewed as the linear expansion of the matrix $K_{\lambda\mu}$ over the set of toric matrices $W_{z0}$, where the coefficients of this expansion are the non-negative integers $(W_{0z})_{\lambda\mu}$. The number $d_O$ is the dimension of the Ocneanu quantum symmetry algebra, it is evaluated by $d_O=Tr(\mathcal{M}\mathcal{M}^{\dagger})$. The algebra of quantum symmetries comes with a  basis (call its
elements $z$) which is special because structure constants of the
algebra, in this basis, are non-negative integers. We introduce the linear map from the space of quantum symmetries
to the space of $d_I \times d_I$ matrices defined by $z \mapsto W_{z0}$. 
This map is not necessarily injective: although elements $z$ of the quantum symmetries are linearly independent, it may not be so for the toric matrices $W_{z0}$ (in particular two distinct elements of the quantum symmetries can sometimes be associated with the same toric matrix).
Let us call $r$ the number of linearly independent matrices $W_{z0}$. 
Equation (\ref{Kmse}) tells us that each  $K_{\lambda\mu}$ (a matrix), defined by (\ref{klm}), can be decomposed on the $r$ dimensional vector space spanned by the vectors (matrices) $W_{z0}$. The number $r$ can be obtained as follows. From (\ref{klm}) we build a matrix K with elements of the form $\textrm{K}_{\{\lambda\mu\},\{\lambda'\mu'\}}$, which means that each line of K is a flattened\footnote{By flattened matrix we mean that if $K_{\lambda\mu} = \left( \begin{array}{ccc} a & .. & b \\ .. & .. & .. \\ c & .. &  d \end{array}\right)$, then the flattened matrix is $(a \, .. \, b \, .. \, .. \, .. \, c\, .. \, d )$.} matrix $K_{\lambda\mu}$. Then $r$ is obtained as the (line) rank of the matrix K, since the rank gives precisely the maximal number of independent lines of K, therefore the number $r$ of linearly independent matrices $W_{z0}$.
Two cases are therefore to be considered: depending if toric matrices are all linearly independent (the map $z \mapsto W_{z0}$ is injective and $r=d_O$) or not $(r<d_O)$. \\

We also introduce a scalar product in the vector space of quantum
symmetries for which the $z$ basis is orthonormal. We consider the
squared norm of the element $ \sum_z (W_{0z})_{\lambda \mu}  z $ and
denote it $\vert \vert  K_{\lambda \mu} \vert \vert^2 $. This is an
abuse of notation, ``justified'' by equation (\ref{Kmse}), and in
the same way, we shall often talk, in what follows,  of the ``squared
norm of the matrix $K_{\lambda \mu}$'', therefore identifying $z$ with
$W_{z0}$, although the linear map is not necessarily an isomorphism.
We have the following property:
\begin{prop}
The squared norm of the matrix $K_{\lambda\mu}$ is given by:
\begin{equation}
||K_{\lambda\mu}||^2 = (K_{\lambda\mu})_{\lambda^*\mu^*} \;.
\label{snorm}
\end{equation}
\end{prop}
{\bf Proof}: We have:
\begin{eqnarray*}
||K_{\lambda\mu}||^2 &=& \sum_{z}\left|(W_{0z})_{\lambda,\mu}\right|^2 \\
{ }                  &=& \sum_z (W_{0z})_{\lambda\mu} \, (W_{z0})_{\lambda^*\mu^*} \\
{ }                   &=& (K_{\lambda\mu})_{\lambda^*\mu^*}
\end{eqnarray*}
From the first to the second line we used the following property:
\begin{equation} 
(W_{0z})_{\lambda\mu}=(W_{z0})_{\lambda^*\mu^*}
\label{Wz0W0z}
\end{equation} 
that can be derived from the relation $V_{\lambda^*\mu^*}=(V_{\lambda\mu})^{tr}$, where $\lambda^*$ is the conjugated irrep of $\lambda$ (see \cite{Pet_Zub-Occells}). From the second to the third line we use Eq. (\ref{Kmse}) in matrix components.\hfill $\blacksquare$\\

We now treat the two cases to be considered. Note: an explicit study of all cases seems to indicate that the linear independence (or not) of the toric matrices  reflects  the commutativity (or not)  of the quantum symmetry algebra.

\paragraph{Non-degenerate case $r=d_O$.}
This happens when all toric matrices $W_{z0}$ are linearly independent.   
The set of $K_{\lambda\mu}$ matrices are calculated from the initial data $\mathcal{M}$ and $N_{\lambda}$ from (\ref{klm}). 
The determination of the toric matrices $W_{z0}$ are recursively obtained from a discussion of the squared norm of matrices $K_{\lambda\mu}$, directly obtained from (\ref{snorm}), which has to be a sum of squared  integers.

\begin{itemize}
\item Consider the set of linearly independent matrices $K_{\lambda\mu}$ of squared norm 1. From (\ref{Kmse}) the solution is that each such matrix is equal to a toric matrix $W_{z0}$.

\item Next we consider the set of linearly independent matrices $K_{\lambda\mu}$ of squared norm 2. In this case from (\ref{Kmse}) each such matrix is equal to the sum of two toric matrices. We have three cases: (i) $K_{\lambda\mu}$ is equal to the sum of two already determined toric matrices (no new information); (ii) it is the sum of an already determined toric matrix and of a new one; (iii) it is equal to the sum of two new toric matrices. To distinguish from cases (ii) and (iii), we calculate the set of
differences $K_{\lambda\mu} - W_i$ where $W_i$ runs into the set of determined toric matrices, and check if the obtained matrix has non-negative integer coefficients: in this case we determine a new toric matrix given by $K_{\lambda\mu} - W_i$.   
 
\item Next we consider the set of linearly independent matrices $K_{\lambda\mu}$ of squared norm 3. From (\ref{Kmse}) each such matrix is equal to the sum of three toric matrices. Either (i) $K_{\lambda\mu}$ is equal to the sum of three already determined toric matrices; (ii) it is equal to the sum of a determined toric matrix and of two new ones; (iii) it is equal to the sum of two already determined matrices and a new one; or (iv) it is equal to the sum of three new toric matrices.
We calculate the set of differences $K_{\lambda\mu} - W_i$ and $K_{\lambda\mu} - W_{i}- W_{j}$ where $W_i, W_j$ runs into the set of determined toric matrices, and check whenever the obtained matrix has non-negative integer coefficients.

\item For the set of linearly independent matrices $K_{\lambda\mu}$ of squared norm 4 there are two possibilities.  
Either $K_{\lambda\mu}$ is the sum of four toric matrices, either it is equal to twice a toric matrix. In the last case, the matrix elements of $K_{\lambda\mu}$ should be either 0 or a multiple of 2, and the new toric matrix is obtained as $K_{\lambda\mu}/2$. If not, a similar discussion as the one made for the previous items allows the determination of the new toric matrices.

\item The next step is to generalize the previous discussions for higher values of the squared norm, in a straightforward way.
\end{itemize}

Once the set of toric matrices $W_{z0}$ is determined, we can of course use equation (\ref{MSE}) to check the results.

\paragraph{Degenerate case $r<d_O$.}
The integer $r$ may be
strictly smaller than $d_O$: this happens when toric matrices $W_{z0}$ are {\sl not} linearly independent.  In order to better illustrate what has to be done in this case, let us treat a ``virtual'' example. Suppose the dimension of the Ocneanu algebra is $d_O=3$, and call $z_1,z_2,z_3$ the basis elements. The corresponding toric matrices are $W_{z_1},W_{z_2},W_{z_3}$, and suppose they are not linearly independent. For example let us take $W_{z_3} = W_{z_1} + W_{z_2}$, in this case we have $r=2<d_O$. 
We still use the same scalar product in the algebra of quantum
symmetries, and  the norm of $z_3$ is of course $1$, but, because of
the abuse of langage and notation already made before, we shall say
that the ``squared norm'' of $W_{z_3}$ is equal to 1 (and not 2, of
course!). The problem arising from the fact that toric matrices may not be linearly independent, so that the linear expansion (\ref{Kmse}) of $K_{\lambda\mu}$ over the family of toric matrices may be not unique, can be solved by considering the squared norm of $K_{\lambda\mu}$. Continuing with our virtual example, we could hesitate between writing $K_{\lambda\mu} = W_{z_1} + 2 W_{z_2}$ or $K_{\lambda\mu} = W_{z_2} + W_{z_3}$, since $W_{z_3}= W_{z_1} + W_{z_2}$. In the first case the corresponding squared norm would be 5, and in the second case it would be 2. In all cases we have met, the knowledge of the squared norm of $K_{\lambda\mu}$ from equation (\ref{snorm}) is sufficient to bypass the ambiguity and obtain the correct linear expansion. The determination of the toric matrices can then be done step by step, in the same way as we did in the non degenerate case, starting from squared norm 1 to higher values.  We refer to the $\widehat{su}(3)$ case at level 9 treated in the next section for more technical details.

\subsection{Determination of double toric matrices $W_{xy}$}
Once we have determined the toric matrices $W_{x0}$, we calculate the following set of matrices:
\begin{equation}
K_{\lambda\mu}^x = N_{\lambda}\,W_{x0}\,N_{\mu}^{tr}
\end{equation}
Then equation (\ref{MSEgen}) reads:
\begin{equation}
K_{\lambda\mu}^x = \sum_z (W_{xz})_{\lambda\mu} W_{z0} \;.
\label{klmx}
\end{equation}
This equation can be viewed as the linear expansion of the matrix $K_{\lambda\mu}^x$ over the set of toric matrices $W_{z0}$, where the coefficients of this expansion are the non-negative integers $(W_{xz})_{\lambda\mu}$, that we want to determine. In the non degenerate case, toric matrices $W_{z0}$ are linearly independent, the decomposition (\ref{klmx}) is unique and the calculation is straightforward. In the degenerate case, some care has to be taken since toric matrices $W_{z0}$ are not linearly independent: the expansion (\ref{klmx}) is therefore not unique. Some coefficients may remain free and one needs further information to a complete determination (see next subsection).\\

The coefficients $(W_{xz})_{\lambda\mu}$ can also be encoded in the double fusion matrices $V_{\lambda\mu}$, that satisfy the double fusion equations (\ref{dfe}). Setting 
$\mu=\mu'=0$, $\lambda=\lambda'=0$ and $\lambda'=\mu=0$ respectively in Eq. (\ref{dfe}) gives:
\begin{eqnarray}
\label{Vfusion1}
V_{\lambda 0} \, V_{\lambda' 0} &=& \sum_{\lambda''} N_{\lambda\lambda'}^{\lambda''} \, V_{\lambda'' 0} \;, \\
\label{Vfusion2}
V_{0 \mu} \, V_{0 \mu'} &=& \sum_{\mu''} N_{\mu\mu'}^{\mu''} \, V_{0\mu''} \;, \\
\label{Vfusion3}
V_{\lambda\mu'} &=& V_{\lambda 0} \, V_{0 \mu'}  = V_{0\mu'} \, V_{\lambda 0} \;.
\end{eqnarray}
From Eqs.(\ref{Vfusion1}) and (\ref{Vfusion2}), we see that the set of matrices $V_{\lambda0}$ and $V_{0\lambda}$ satisfy the fusion algebra. These matrices can therefore be determined using these equations from the subset of matrices $V_{f0}$ and $V_{0f}$, where $f$ stands for the fundamental generators of the fusion algebra. For $\widehat{su}(2)$, there is one generator $f=1$, while for $\widehat{su}(3)$, there are two conjugated generators $(1,0)$ and $(0,1)$. The determination of double fusion matrices is reduced, by the use of Eqs. (\ref{Vfusion1}--\ref{Vfusion3}), to the determination of the generators $V_{f0}$ and $V_{0f}$. It is therefore sufficient to solve
Eq. (\ref{klmx}) only for the pair of indices $(\lambda\mu)=(f0)$ and $(\lambda\mu)=(0f)$, and then use Eqs. (\ref{Vfusion1}--\ref{Vfusion3}), which simplifies a lot the computational task.

\subsection{Determination of the Ocneanu algebra $O_{x}$}
The matrices $V_{f0}$ and $V_{0f}$ are the adjacency matrices of the Ocneanu graph. We denote $O_{f_L} = V_{f0}$ and $O_{f_R}=V_{0f}$, where $f_L$ and $f_R$ now stands for the left and right generators of the Ocneanu quantum symmetry algebra.
For $SU(n)$, there are $n-1$ generators $f$ of the fusion algebra, and therefore $2(n-1)$ generators of the quantum symmetry algebra. The Ocneanu graph is also the Cayley graph of multiplication by these generators. From the multiplication by these generators, we can reconstruct the full table of multiplication of the quantum symmetry algebra (with elements denoted $x,y,z$)
\begin{equation}
x \, y = \sum_z \mathcal{O}_{xy}^z \, z \;.
\end{equation}
This multiplication table is encoded in the ``quantum symmetry matrices'' $O_x$, which are the graph algebra matrices of the Ocneanu graph, with coefficients $(O_x)_{yz} = \mathcal{O}_{xy}^z$. They satisfy the following relations (take care with the order of indices since the quantum symmetry algebra may be non commutative):
\begin{equation}
O_x \, O_y = \sum_z (O_y)_{xz} \, O_z \;.
\label{ox}
\end{equation}
Once the generators $O_{f_L} = V_{f0}$ and $O_{f_R}=V_{0f}$ have been determined from the previous step, all quantum symmetry matrices can be computed from (\ref{ox}).\\

In the degenerate case the determination of the double toric matrices $W_{xy}$ from equation (\ref{klmx}) is not straightforward, some coefficients being still free. A solution to this problem is provided by an analysis of the structure of the Ocneanu graph itself, since it must satisfy some conjugation and chiral conjugation properties (we refer to the level 9 $\widehat{su}(3)$ example treated in the next section for further details). Further compatibility conditions have also to be satisfied and can be used to check the results, or to determine the remaining coefficients (for degenerate cases). One of these conditions read \cite{Pet_Zub-Occells,Coq-maroc}:
\begin{equation}
O_x \, V_{\lambda \mu} = V_{\lambda \mu} \, O_x = \sum_z (V_{\lambda \mu})_{xz} \, O_z \;.
\label{OVrel}
\end{equation}
A special case of this equation, for $x=0$, being:
\begin{equation}
W_{yy'} = \sum_z (O_z)_{yy'} \, W_{0z} \;.
\end{equation}

\subsection{Determination of the higher ADE graph $G$}
For any $\widehat{su}(n)$ at level $k$,  we have the
infinite series of $\mathcal{A}_k$ graphs which are the truncated
Weyl alcoves at level $k$ of $SU(n)$ irreps.  Other infinite series are obtained
by orbifolding ($\mathcal{D}_k=\mathcal{A}_k/p$) and conjugation ($
\mathcal{A}_k^*, \mathcal{D}_k^*$) methods, but there are also some
exceptional graphs that can not be obtained in that way. Even
using the fact that graphs have to obey a list of requirements (such
as conjugation, N-ality, spectral properties and that $G$ must be 
an $\mathcal{A}_k$ module), one still needed to use some good  ``computer aided flair'' to find
them \cite{DiFZuber, Pet_Zub-graphs}. The basic method to obtain the exceptional graphs was to use the 
Xu algorithm (see \cite{Xu, Pet_Zub-bound}) for solving (\ref{fusF}), at least when the initial data 
$(F_{\lambda})_{00}$ is known (from conformal embedding for instance).

In this ``historical approach'', the problem of determining the
algebra of quantum symmetries $Oc(G)$ was not addressed and this
algebra was even less used as a tool to determine $G$ itself.
The procedure described in this paper is different. Starting from the modular
invariant $\mathcal{M}_{\lambda\mu} = (V_{\lambda\mu})_{00}$ as initial data, one 
solves the modular splitting equation derived from (\ref{dfe}) (as explained
in the previous section) and determines directly the algebra of
quantum symmetries $Oc(G)$, without knowing what $G$ itself can be.
Then one uses the fact that $G$ should be {\sl both} an $\mathcal{A}_k$
module {\sl and} an $Oc(G)$ module (see comments in \cite{Coq-maroc}).
Denoting $\lambda$ an element of the fusion algebra, the first
module property reads $\lambda \, a = \sum_b (F_{\lambda})_{ab} \, b
$, with coefficients encoded by the annular matrices $F_{\lambda}$.
The associativity property $
(\lambda \, \mu) \, a = \lambda \, (\mu \, a)$ imposes the annular
matrices to satisfy the fusion algebra (\ref{fusF}).
Denoting $x$ an element of the quantum symmetry algebra, the second
module property reads $x \, a = \sum_b (S_{x})_{ab} \, b$, with
coefficients encoded by the dual annular matrices $S_{x}$. The
associativity property $ (x \, y) \, a = x \, (y \, a)$ imposes the
dual annular matrices to satisfy the quantum symmetry algebra (\ref{dualS}).
In some cases (including all Type I cases), $G$ directly appears as a subgraph of the Ocneanu graph. 
In other cases, it appears as a module over the algebra of a particular subgraph.\\

The methods we have described allow for the determination of the
graph $G$ even when orbifold and conjugation arguments from the $\mathcal{A}_k$ graphs do not apply
(the exceptional cases).  It can be used for a general affine algebra
$\widehat{g}_k$ at any given level $k$, once the corresponding
modular invariant is known.
In the next section, we present and
illustrate this method using several exceptional examples. 
In the $su(3)$ family, there are three exceptional graphs with self
fusion. They are called $\mathcal{E}_5, \mathcal{E}_9$ and $\mathcal{E}_{21}$.
In this paper we have chosen $\mathcal{E}_5$ (a kind of generalization of the $E_6$ case of $su(2)$) 
and $\mathcal{E}_9$. The case of $\mathcal{E}_{21}$ (a kind of generalization of the $E_8$ case of $su(2)$) is actually
very simple to discuss, even simpler than $\mathcal{E}_5$  because it does not
admit any non trivial module graph, and we could have described it as
well,  along the same lines. Results concerning $\mathcal{E}_{21}$ and its quantum
symmetries can be found in \cite{Coq_Gil-Tmod, Gil-tesis} (in those
references, the graph itself is a priori given). The $su(3)$ - analogue
of the $E_7$ case of $su(2)$, which is an exceptional twist of $\mathcal{D}_{9}$, can also
be analysed thank's to the modular splitting formula, of course, but
the discussion is quite involved (see \cite{Dah-tesis,Gil-maroc2}).
We refer to \cite{Gil-e4} for a description of an $\widehat{su}(4)$
example. In \cite{Coq_esteban}, these methods were applied to a non
simply-laced example of the $su(2)$ family, where the initial partition
function is not modular invariant (it is invariant under a particular
congruence subgroup) and where there is  no associated quantum groupo\"id.

\subsection{Comments}
 All module, associativity and compatibility conditions described here between the different set of matrices follow from properties of the quantum groupo\"id $\mathcal{B}(G)$ constructed from the higher ADE graph $G$ \cite{Oc-paths,Pet_Zub-Occells,Gil-tesis}. General results have been published on this quantum groupo\"id (see \cite{Oc-paths,Coq_Trinchero,Coq-q6j,kir-ost,Ost}). But we are not aware of any definite list of properties that the graphs $G$ should satisfy to obtain the right classification. The strategy adopted here is to take as granted the existence of a quantum groupo\"{\i}d and its corresponding set of properties, and to derive the graph $G$ as a by-product of the calculations, starting from the only knowledge of the modular invariant. Notice that this seems to be the method adopted by Ocneanu in order to produce his list of $SU(3)$ and $SU(4)$ graphs presented in \cite{Oc-Bariloche}.
One crucial check for the existence of the underlying quantum groupo\"id is the existence of dimensional rules:
\begin{equation}
\dim(\mathcal{B}(G)) = \sum_{\lambda} d_{\lambda}^2 = \sum_x d_x^2 \;,
\end{equation}
where the dimensions $d_{\lambda}$ and $d_x$ are calculated from the
annular and dual annular matrices: $d_{\lambda} = \sum_{a,b}
(F_{\lambda})_{ab}$, $d_x = \sum_{a,b} (S_x)_{ab}$.\\


\section{Examples}

\subsection{The $\mathcal{E}_5$ case of $\widehat{su}(3)$}
\begin{figure}[hhh] 
\begin{center} 
\unitlength 0.30mm 

\begin{picture}(200,180)(0,-10)

\put(0,0){\begin{picture}(40,40)
\put(0,0){\color{green} \circle*{7}} 
\put(40,0){\color{blue} \circle*{7}}
\put(20,30){\color{red} \circle*{7}} 
\put(0,0){\vector(1,0){21}} 
\put(20,0){\line(1,0){20}} 
\put(40,0){\vector(-2,3){11.5}} 
\put(30,15){\line(-2,3){10}} 
\put(20,30){\vector(-2,-3){11.5}} 
\put(10,15){\line(-2,-3){10}}\end{picture}}

\put(40,0){\begin{picture}(40,40) 
\put(40,0){\color{red} \circle*{7}} 
\put(20,30){\color{green} \circle*{7}}
\put(0,0){\vector(1,0){21}} 
\put(20,0){\line(1,0){20}} 
\put(40,0){\vector(-2,3){11.5}} 
\put(30,15){\line(-2,3){10}} 
\put(20,30){\vector(-2,-3){11.5}} 
\put(10,15){\line(-2,-3){10}}\end{picture}} 
 
\put(80,0){\begin{picture}(40,40) 
\put(40,0){\color{green} \circle*{7}}
\put(20,30){\color{blue} \circle*{7}}
\put(0,0){\vector(1,0){21}} 
\put(20,0){\line(1,0){20}} 
\put(40,0){\vector(-2,3){11.5}} 
\put(30,15){\line(-2,3){10}} 
\put(20,30){\vector(-2,-3){11.5}} 
\put(10,15){\line(-2,-3){10}}\end{picture}}

\put(120,0){\begin{picture}(40,40) 
\put(40,0){\color{blue} \circle*{7}}
\put(20,30){\color{red} \circle*{7}} 
\put(0,0){\vector(1,0){21}} 
\put(20,0){\line(1,0){20}} 
\put(40,0){\vector(-2,3){11.5}} 
\put(30,15){\line(-2,3){10}} 
\put(20,30){\vector(-2,-3){11.5}} 
\put(10,15){\line(-2,-3){10}}\end{picture}}

\put(160,0){\begin{picture}(40,40) 
\put(40,0){\color{red} \circle*{7}} 
\put(20,30){\color{green} \circle*{7}}
\put(0,0){\vector(1,0){21}} 
\put(20,0){\line(1,0){20}} 
\put(40,0){\vector(-2,3){11.5}} 
\put(30,15){\line(-2,3){10}} 
\put(20,30){\vector(-2,-3){11.5}} 
\put(10,15){\line(-2,-3){10}}\end{picture}}

\put(20,30){\begin{picture}(40,40) 
\put(20,30){\color{blue} \circle*{7}}
\put(0,0){\vector(1,0){21}} 
\put(20,0){\line(1,0){20}} 
\put(40,0){\vector(-2,3){11.5}} 
\put(30,15){\line(-2,3){10}} 
\put(20,30){\vector(-2,-3){11.5}} 
\put(10,15){\line(-2,-3){10}}\end{picture}} 
 
\put(60,30){\begin{picture}(40,40) 
\put(20,30){\color{red} \circle*{7}} 
\put(0,0){\vector(1,0){21}} 
\put(20,0){\line(1,0){20}} 
\put(40,0){\vector(-2,3){11.5}} 
\put(30,15){\line(-2,3){10}} 
\put(20,30){\vector(-2,-3){11.5}} 
\put(10,15){\line(-2,-3){10}}\end{picture}} 
 
\put(100,30){\begin{picture}(40,40) 
\put(20,30){\color{green} \circle*{7}}
\put(0,0){\vector(1,0){21}} 
\put(20,0){\line(1,0){20}} 
\put(40,0){\vector(-2,3){11.5}} 
\put(30,15){\line(-2,3){10}} 
\put(20,30){\vector(-2,-3){11.5}} 
\put(10,15){\line(-2,-3){10}}\end{picture}}

\put(140,30){\begin{picture}(40,40) 
\put(20,30){\color{blue} \circle*{7}} 
\put(0,0){\vector(1,0){21}} 
\put(20,0){\line(1,0){20}} 
\put(40,0){\vector(-2,3){11.5}} 
\put(30,15){\line(-2,3){10}} 
\put(20,30){\vector(-2,-3){11.5}} 
\put(10,15){\line(-2,-3){10}}\end{picture}}

\put(40,60){\begin{picture}(40,40) 
\put(20,30){\color{green} \circle*{7}}
\put(0,0){\vector(1,0){21}} 
\put(20,0){\line(1,0){20}} 
\put(40,0){\vector(-2,3){11.5}} 
\put(30,15){\line(-2,3){10}} 
\put(20,30){\vector(-2,-3){11.5}} 
\put(10,15){\line(-2,-3){10}}\end{picture}} 
 
\put(80,60){\begin{picture}(40,40) 
\put(20,30){\color{blue} \circle*{7}}
\put(0,0){\vector(1,0){21}} 
\put(20,0){\line(1,0){20}} 
\put(40,0){\vector(-2,3){11.5}} 
\put(30,15){\line(-2,3){10}} 
\put(20,30){\vector(-2,-3){11.5}} 
\put(10,15){\line(-2,-3){10}}\end{picture}}

\put(120,60){\begin{picture}(40,40) 
\put(20,30){\color{red} \circle*{7}} 
\put(0,0){\vector(1,0){21}} 
\put(20,0){\line(1,0){20}} 
\put(40,0){\vector(-2,3){11.5}} 
\put(30,15){\line(-2,3){10}} 
\put(20,30){\vector(-2,-3){11.5}} 
\put(10,15){\line(-2,-3){10}}\end{picture}}

\put(60,90){\begin{picture}(40,40) 
\put(20,30){\color{red} \circle*{7}} 
\put(0,0){\vector(1,0){21}} 
\put(20,0){\line(1,0){20}} 
\put(40,0){\vector(-2,3){11.5}} 
\put(30,15){\line(-2,3){10}} 
\put(20,30){\vector(-2,-3){11.5}} 
\put(10,15){\line(-2,-3){10}}\end{picture}}

\put(100,90){\begin{picture}(40,40) 
\put(20,30){\color{green} \circle*{7}}
\put(0,0){\vector(1,0){21}} 
\put(20,0){\line(1,0){20}} 
\put(40,0){\vector(-2,3){11.5}} 
\put(30,15){\line(-2,3){10}} 
\put(20,30){\vector(-2,-3){11.5}}
\put(10,15){\line(-2,-3){10}}\end{picture}}

\put(80,120){\begin{picture}(40,40) 
\put(20,30){\color{blue} \circle*{7}}
\put(0,0){\vector(1,0){21}} 
\put(20,0){\line(1,0){20}} 
\put(40,0){\vector(-2,3){11.5}} 
\put(30,15){\line(-2,3){10}} 
\put(20,30){\vector(-2,-3){11.5}} 
\put(10,15){\line(-2,-3){10}}\end{picture}} 
 
\put(-5,-10){\makebox(0,0){(0,0)}} 
\put(40,-10){\makebox(0,0){(1,0)}} 
\put(3,30){\makebox(0,0){(0,1)}} 
\put(200,-10){\makebox(0,0){$(5,0)$}} 
\put(100,160){\makebox(0,0){$(0,5)$}} 
\end{picture}
\end{center} 
\caption{The ${\cal A}_5$ diagram.} 
\label{graphA5} 
\end{figure}
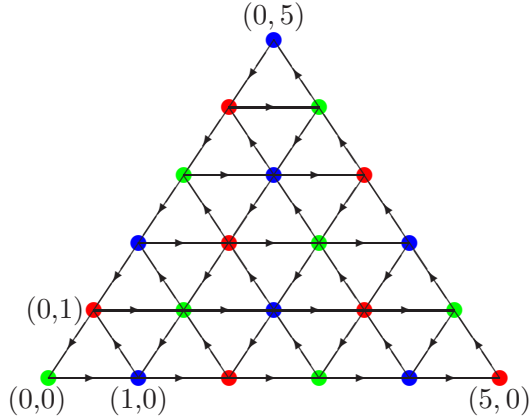

\begin{figure}[hhh] 
\begin{center} 
\unitlength 0.25mm 
 
\begin{picture}(400,410) 

\put(20,20){\begin{picture}(360,370)

\qbezier[60](0,325)(-160,180)(90,190)
\qbezier[10](0,235)(15,250)(0,260)
\qbezier[10](0,235)(-15,250)(0,260)
\qbezier[100](90,370)(350,450)(180,235)
\qbezier[10](180,325)(195,340)(180,350)
\qbezier[10](180,325)(165,340)(180,350)

\qbezier[10](240,135)(255,150)(240,160)
\qbezier[10](240,135)(225,150)(240,160)
\qbezier[30](300,135)(280,80)(210,90)
\qbezier[30](330,90)(260,100)(240,45)
\qbezier[10](300,45)(315,30)(300,20)
\qbezier[10](300,45)(285,30)(300,20)

\qbezier[100](60,235)(45,5)(270,0)
\qbezier[140](30,280)(-20,-120)(360,45)
\qbezier[80](120,235)(90,120)(180,45)

\qbezier[130](60,325)(350,490)(360,135)
\qbezier[140](120,325)(300,440)(270,190)


\put(0,190){\begin{picture}(180,180)

\put(0,45){\begin{picture}(60,45) 
\put(0,0){\color{green} \circle*{7}} 
\put(60,0){\color{blue} \circle*{7}}
\put(30,45){\color{red} \circle*{7}} 
\put(0,0){\vector(1,0){32.5}} 
\put(30,0){\line(1,0){30}} 
\put(60,0){\vector(-2,3){16.5}} 
\put(45,22.5){\line(-2,3){15}} 
\put(30,45){\vector(-2,-3){16.5}} 
\put(15,22.5){\line(-2,-3){15}} 
\end{picture}} 
 
\put(120,45){\begin{picture}(60,45) 
\put(0,0){\color{green} \circle*{7}} 
\put(60,0){\color{blue} \circle*{7}}
\put(30,45){\color{red} \circle*{7}} 
\put(0,0){\vector(1,0){32.5}} 
\put(30,0){\line(1,0){30}} 
\put(60,0){\vector(-2,3){16.5}} 
\put(45,22.5){\line(-2,3){15}} 
\put(30,45){\vector(-2,-3){16.5}} 
\put(15,22.5){\line(-2,-3){15}} 
\end{picture}} 
 
\put(60,135){\begin{picture}(60,45) 
\put(0,0){\color{green} \circle*{7}} 
\put(60,0){\color{blue} \circle*{7}}
\put(30,45){\color{red} \circle*{7}} 
\put(0,0){\vector(1,0){32.5}} 
\put(30,0){\line(1,0){30}} 
\put(60,0){\vector(-2,3){16.5}} 
\put(45,22.5){\line(-2,3){15}} 
\put(30,45){\vector(-2,-3){16.5}} 
\put(15,22.5){\line(-2,-3){15}} 
\end{picture}}

\put(0,90){\begin{picture}(60,45) 
\put(0,45){\color{blue} \circle*{7}}
\put(60,45){\vector(-1,0){32.5}} 
\put(30,45){\line(-1,0){30}} 
\put(30,0){\vector(2,3){16.5}} 
\put(45,22.5){\line(2,3){15}} 
\put(0,45){\vector(2,-3){16.5}} 
\put(15,22.5){\line(2,-3){15}} 
\end{picture}} 
 
\put(120,90){\begin{picture}(60,45) 
\put(60,45){\color{green} \circle*{7}} 
\put(60,45){\vector(-1,0){32.5}} 
\put(30,45){\line(-1,0){30}} 
\put(30,0){\vector(2,3){16.5}} 
\put(45,22.5){\line(2,3){15}} 
\put(0,45){\vector(2,-3){16.5}} 
\put(15,22.5){\line(2,-3){15}} 
\end{picture}} 
 
\put(60,0){\begin{picture}(60,45) 
\put(30,0){\color{red} \circle*{7}} 
\put(60,45){\vector(-1,0){32.5}} 
\put(30,45){\line(-1,0){30}} 
\put(30,0){\vector(2,3){16.5}} 
\put(45,22.5){\line(2,3){15}} 
\put(0,45){\vector(2,-3){16.5}} 
\put(15,22.5){\line(2,-3){15}} 
\end{picture}} 
 
\put(60,135){\vector(0,-1){47.5}} 
\put(60,90){\line(0,-1){45}} 
\put(60,45){\vector(2,1){47.2}} 
\put(105,67.5){\line(2,1){45}} 
\put(150,90){\vector(-2,1){47.2}} 
\put(105,112.5){\line(-2,1){45}} 
 
\put(120,45){\vector(0,1){47.5}} 
\put(120,90){\line(0,1){45}} 
\put(120,135){\vector(-2,-1){47.2}} 
\put(75,112.5){\line(-2,-1){45}} 
\put(30,90){\vector(2,-1){47.2}} 
\put(75,67.5){\line(2,-1){45}}

\put(-30,47){\makebox(0,0){\scriptsize{$1_0\otimes 1_0$}}} 
\put(210,45){\makebox(0,0){\scriptsize{$1_4\otimes 1_0$}}} 
\put(-30,135){\makebox(0,0){\scriptsize{$1_1\otimes 1_0$}}} 
\put(210,135){\makebox(0,0){\scriptsize{$1_3\otimes 1_0$}}} 
\put(90,-12){\makebox(0,0){\scriptsize{$1_5\otimes 1_0$}}} 
\put(60,180){\makebox(0,0){\scriptsize{$1_2\otimes 1_0$}}} 

\put(45,35){\makebox(0,0){\scriptsize{$2_1\otimes 1_0$}}}  
\put(137,35){\makebox(0,0){\scriptsize{$2_0\otimes 1_0$}}} 
\put(43,148){\makebox(0,0){\scriptsize{$2_3\otimes 1_0$}}} 
\put(136,148){\makebox(0,0){\scriptsize{$2_4\otimes 1_0$}}} 
\put(5,92){\makebox(0,0){\scriptsize{$2_2\otimes 1_0$}}} 
\put(177,92){\makebox(0,0){\scriptsize{$2_5\otimes 1_0$}}} 

\end{picture}}


\put(180,0){\begin{picture}(180,180)

\put(0,45){\begin{picture}(60,45) 
\put(0,0){\color{green} \circle*{7}} 
\put(60,0){\color{blue} \circle*{7}}
\put(30,45){\color{red} \circle*{7}} 
\put(0,0){\vector(1,0){32.5}} 
\put(30,0){\line(1,0){30}} 
\put(60,0){\vector(-2,3){16.5}} 
\put(45,22.5){\line(-2,3){15}} 
\put(30,45){\vector(-2,-3){16.5}} 
\put(15,22.5){\line(-2,-3){15}} 
\end{picture}} 
 
\put(120,45){\begin{picture}(60,45) 
\put(0,0){\color{green} \circle*{7}}  
\put(60,0){\color{blue} \circle*{7}}
\put(30,45){\color{red} \circle*{7}} 
\put(0,0){\vector(1,0){32.5}} 
\put(30,0){\line(1,0){30}} 
\put(60,0){\vector(-2,3){16.5}} 
\put(45,22.5){\line(-2,3){15}} 
\put(30,45){\vector(-2,-3){16.5}} 
\put(15,22.5){\line(-2,-3){15}} 
\end{picture}} 
 
\put(60,135){\begin{picture}(60,45) 
\put(0,0){\color{green} \circle*{7}} 
\put(60,0){\color{blue} \circle*{7}}
\put(30,45){\color{red} \circle*{7}} 
\put(0,0){\vector(1,0){32.5}} 
\put(30,0){\line(1,0){30}} 
\put(60,0){\vector(-2,3){16.5}} 
\put(45,22.5){\line(-2,3){15}} 
\put(30,45){\vector(-2,-3){16.5}} 
\put(15,22.5){\line(-2,-3){15}} 
\end{picture}}

\put(0,90){\begin{picture}(60,45) 
\put(0,45){\color{blue} \circle*{7}}\put(60,45){\vector(-1,0){32.5}} 
\put(30,45){\line(-1,0){30}} 
\put(30,0){\vector(2,3){16.5}} 
\put(45,22.5){\line(2,3){15}} 
\put(0,45){\vector(2,-3){16.5}} 
\put(15,22.5){\line(2,-3){15}} 
\end{picture}} 
 
\put(120,90){\begin{picture}(60,45) 
\put(60,45){\color{green} \circle*{7}} 
\put(60,45){\vector(-1,0){32.5}} 
\put(30,45){\line(-1,0){30}} 
\put(30,0){\vector(2,3){16.5}} 
\put(45,22.5){\line(2,3){15}} 
\put(0,45){\vector(2,-3){16.5}} 
\put(15,22.5){\line(2,-3){15}} 
\end{picture}} 
 
\put(60,0){\begin{picture}(60,45) 
\put(30,0){\color{red} \circle*{7}} 
\put(60,45){\vector(-1,0){32.5}} 
\put(30,45){\line(-1,0){30}} 
\put(30,0){\vector(2,3){16.5}} 
\put(45,22.5){\line(2,3){15}} 
\put(0,45){\vector(2,-3){16.5}} 
\put(15,22.5){\line(2,-3){15}} 
\end{picture}} 
 
\put(60,135){\vector(0,-1){47.5}} 
\put(60,90){\line(0,-1){45}} 
\put(60,45){\vector(2,1){47.2}} 
\put(105,67.5){\line(2,1){45}} 
\put(150,90){\vector(-2,1){47.2}} 
\put(105,112.5){\line(-2,1){45}} 
 
\put(120,45){\vector(0,1){47.5}} 
\put(120,90){\line(0,1){45}} 
\put(120,135){\vector(-2,-1){47.2}} 
\put(75,112.5){\line(-2,-1){45}} 
\put(30,90){\vector(2,-1){47.2}} 
\put(75,67.5){\line(2,-1){45}}

\put(-30,47){\makebox(0,0){\scriptsize{$1_0\otimes 2_0$}}} 
\put(210,45){\makebox(0,0){\scriptsize{$1_4\otimes 2_0$}}} 
\put(-30,135){\makebox(0,0){\scriptsize{$1_1\otimes 2_0$}}} 
\put(210,135){\makebox(0,0){\scriptsize{$1_3\otimes 2_0$}}} 
\put(90,-12){\makebox(0,0){\scriptsize{$1_5\otimes 2_0$}}} 
\put(90,195){\makebox(0,0){\scriptsize{$1_2\otimes 2_0$}}} 

\put(45,35){\makebox(0,0){\scriptsize{$2_1\otimes 2_0$}}}  
\put(137,35){\makebox(0,0){\scriptsize{$2_0\otimes 2_0$}}} 
\put(43,148){\makebox(0,0){\scriptsize{$2_3\otimes 2_0$}}} 
\put(136,148){\makebox(0,0){\scriptsize{$2_4\otimes 2_0$}}} 
\put(5,92){\makebox(0,0){\scriptsize{$2_2\otimes 2_0$}}} 
\put(177,92){\makebox(0,0){\scriptsize{$2_5\otimes 2_0$}}}

\end{picture}} 

\end{picture}}

\end{picture} 
\end{center} 
\caption{Ocneanu graph $Oc({\cal E}_5)$. The two left chiral generators are $2_1 \otimes 1_0$ and $2_2 \otimes 1_0$, the two right chiral generators are $1_5 \otimes 2_0$ and $1_4 \otimes 2_0$.} 
\label{OcE5} 
\end{figure}
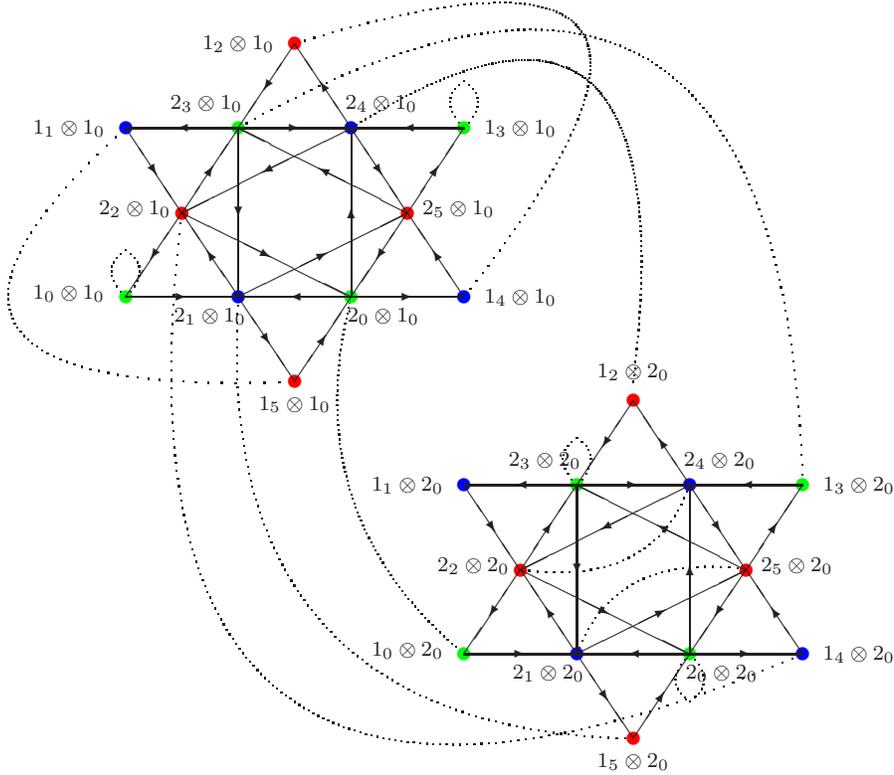

We start with the $\widehat{su}(3)_{5}$ modular invariant partition function:
\begin{eqnarray} 
\nonumber \mathcal{Z} =   &=& |\chi_{(0,0)}^5 + \chi_{(2,2)}^5|^2 + |\chi_{(0,2)}^5 + \chi_{(3,2)}^5|^2 
 + |\chi_{(2,0)}^5 + \chi_{(2,3)}^5|^2 \\ 
{ } &+& |\chi_{(2,1)}^5 + \chi_{(0,5)}^5|^2 + |\chi_{(3,0)}^5 + \chi_{(0,3)}^5|^2 + |\chi_{(1,2)}^5 + \chi_{(5,0)}^5|^2 \;,
\end{eqnarray} 
where $\chi_{\lambda}^5$'s are the characters of $\widehat{su}(3)_{5}$, labelled by $\lambda=(\lambda_1,\lambda_2)$ with $0\leq \lambda_1,\lambda_2\leq 5$, $\lambda_1+\lambda_2\le 5$. The modular invariant matrix $\mathcal{M}$ is read from $\mathcal{Z}$ when the later is written\footnote{Some authors write instead $\mathcal{Z} = \sum_{\lambda}\chi_{\lambda}\,\mathcal{M}_{\lambda\mu^*}\,\bar{\chi}_{\mu}$, and therefore some care has to be taken in order to compare results since conjugated cases  (in particular figures \ref{OcE5} and \ref{OcE5conj}) must then be interchanged. Here we follow the convention made in \cite{Coq-maroc}. } $\mathcal{Z} = \sum_{\lambda}\chi_{\lambda}\,\mathcal{M}_{\lambda\mu}\,\bar{\chi}_{\mu}$. The number of irreps is $d_{\mathcal{A}}= 21$.  
$\lambda=(0,0)$ is the trivial representation and there are two fundamental irreps $(1,0)$ and $(0,1)=(1,0)^*$, where $(\lambda_1,\lambda_2)^*=(\lambda_2,\lambda_1)$ is the conjugated irrep.  $N_{(1,0)}$ is the adjacency matrix of the oriented graph 
$\mathcal{A}_5$, which is the truncated Weyl alcove of SU(3) irreps at level $k=5$ (see figure \ref{graphA5}). The fusion matrix $N_{(0,1)}$ is the transposed matrix of $N_{(1,0)}$ and is the adjacency matrix of the same graph with reversed arrows.
Once $N_{(1,0)}$ is known, the other fusion matrices can be obtained from the {\bf truncated recursion formulae of $SU(3)$ irreps}, applied for increasing level up to $k$:
\begin{eqnarray}
N_{(\lambda,\mu)} &=& N_{(1,0)} \, N_{(\lambda-1,\mu)} - N_{(\lambda-1,\mu-1)} - 
N_{(\lambda-2,\mu+1)} \qquad \qquad \textrm{if } \mu \not= 0 \nonumber \\
N_{(\lambda,0)} &=& N_{(1,0)} \, N_{(\lambda-1,0)} - N_{(\lambda-2,1)} 
\label{su3recform} \\
N_{(0,\lambda)} &=& (N_{(\lambda,0)})^{tr} \nonumber
\label{su3recrel}
\end{eqnarray} 
where it is understood that $N_{(\lambda,\mu)} =0 $ if $\lambda < 0$ or $\mu < 0$.


\paragraph{Determination of toric matrices $W_{z0}$}
We have $d_O = Tr(\mathcal{M}\mathcal{M}^{\dagger}) = 24$. The matrices $K_{\lambda\mu}=N_{\lambda}\,\mathcal{M}\,N_{\mu}^{tr}$ span a vector space of dimension $r=24$. Since $r=d_O$, the toric matrices $W_{x0}$ are linearly independent and form a special basis for this vector space. For each matrix $K_{\lambda\mu}$ we calculate the squared norm given by $||K_{\lambda\mu}||^2=(K_{\lambda\mu})_{\lambda^*\mu^*}$. 
\begin{itemize}
\item For squared norm 1 we have 21 linearly independent matrices $K_{\lambda\mu}$, each one being equal to a toric matrix $W_{z0}$.  
\item There are 45 linearly independent matrices $K_{\lambda\mu}$ of squared norm 2. Some of them are equal to the sum of two already determined toric matrices. For a matrix not satisfying this property, say $K_{ab}$, we build the set of matrices $K_{ab} - W_x$, where $W_x$ runs into the set of determined toric matrices, and look for those which have non-negative integer coefficients.
This condition is strong enough and leads to only one solution (if $K_{ab}$ is the sum of a determined matrix and a new one). We determine in that way the last three toric matrices. 
\item We have therefore determined the set of 24 toric matrices $W_x$, with $0 \leq x \leq 23$ and we can check our result by an explicit verification of the modular splitting equation (\ref{MSE}).
\end{itemize}

\paragraph{Determination of $V_{\lambda\mu}$} 
Having determined the set of toric matrices $W_{x0}$, we compute the set of matrices $K_{\lambda\mu}^x = N_{\lambda}\,W_{x0}\,N_{\mu}^{tr}$. For $SU(3)$ cases, all double fusion matrices $V_{\lambda\mu}$ are generated by the two fundamental matrices $V_{(1,0),(0,0)}$, $V_{(0,0),(1,0)}$ and their transposed $V_{(0,1),(0,0)}=V_{(1,0),(0,0)}^{tr}$, $V_{(0,0),(0,1)}=V_{(0,0),(1,0)}^{tr}$. In order to determine these matrices, it is therefore sufficient to compute the decomposition of $K_{(1,0),(0,0)}^x$ and $K_{(0,0),(1,0)}^x$ on the set of toric matrices $W_{x0}$ using Eq.(\ref{klmx}). The calculation is straightforward. From the knowledge of the fundamental matrices $V_{(1,0),(0,0)}$, $V_{(0,0),(1,0)}$ and their transposed, all double fusion matrices $V_{\lambda\mu}$ are recursively calculated from Eqs.(\ref{Vfusion1}--\ref{Vfusion3}).

\paragraph{The Ocneanu graph of quantum symmetries}
The four fundamental matrices explicitly given below, in Eqs.(\ref{vije5}), are the adjacency matrices of the graph of quantum symmetries (Ocneanu graph) associated to the initial modular invariant. We display in figure \ref{OcE5} the graph corresponding to the matrix $V_{(1,0),(0,0)}$ associated to the vertex labelled by $2_1\otimes1_0$. $V_{(0,0),(1,0)}$ is associated to the vertex $1_5\otimes 2_0$, and instead of displaying the corresponding arrows, we display the action of the chiral conjugation $C$ in order to not clutter the figure (warning: see the last footnote). The arrows corresponding to the matrix $V_{(0,1),(0,0)}$, associated to the vertex $2_2\otimes1_0$, are obtained by reversing the ones of figure \ref{OcE5}; for the matrix $V_{(0,0),(0,1)}$, associated to the vertex $1_4\otimes 2_0$, we use the chiral conjugation and the reversed arrows.

\paragraph{The generalized Dynkin diagram $\mathcal{E}_5$}
The graph of figure \ref{OcE5} is made of two copies of the generalized Dynkin diagram $\mathcal{E}_5$. The ${\mathcal E}_5$ graph has 12 vertices denoted by $1_i, 2_i$, $i =0,1,\ldots,5$. The unit is $1_0$ and the generators are $2_1$ and $2_2$, the orientation  of the graph corresponds to multiplication by $2_1$.
Conjugation corresponds to the symmetry with respect to the axis passing through vertices $1_0$ and $1_3$: 
$1_0^*=1_0, 1_1^*=1_5, 1_2^*=1_4, 1_3^*=1_3$; $2_0^*=2_3, 2_1^*=2_2, 2_4^*=2_5$.  
The $\mathcal{E}_5$ graph determines in a unique way its graph algebra (it is a subgroup graph). The commutative multiplication table is given by:
\begin{equation} 
\begin{array}{rcl} 
1_i . 1_j & = & 1_{i+j}    \\ 
1_i . 2_j = 2_i . 1_j & = & 2_{i+j}  \qquad \qquad \qquad \qquad \qquad \qquad \qquad i,j =0,1,\ldots,5 \; \mod 6 \\ 
2_i . 2_j & = & 2_{i+j} + 2_{i+j-3} + 1_{i+j-3} 
\end{array} 
\label{mult_E5} 
\end{equation} 
From this multiplication table we get the graph algebra matrices $G_a$ associated to the vertices 
$a \in \mathcal{E}_5$. The one corresponding to the generator $2_1$ is the adjacency matrix of the graph. 
The vector space spanned by vertices of $\mathcal{E}_5$ is a module under the action of vertices of ${\cal A}_5$, the action being encoded by the annular matrices $F_{\lambda}$ obtained form the recurrence relation (\ref{su3recform}) with starting point $F_{(0,0)} = \munite_{12}$, $F_{(1,0)}=G_{2_1}$ and $F_{(0,1)}=G_{2_2}$. 

Choosing a special ordering in the set of indices $z$ of the algebra of quantum symmetries, and using the $12\times 12$ graph algebra matrices $G_a$ of the graph $\mathcal{E}_5$, the fundamental double fusion matrices are given by
\begin{equation}
\begin{array}{lcl}
V_{(1,0),(0,0)} = \left(
\begin{array}{c|c}
G_{2_1} & . \\
\hline
. & G_{2_1}
\end{array}
\right) 
&\qquad \qquad& 
V_{(0,0),(1,0)} = \left(
\begin{array}{c|c}
. & G_{1_5} \\
\hline
G_{1_2} & G_{1_2}+G_{1_5}
\end{array}
\right) 
 \\
{} & {} \\
V_{(0,1),(0,0)} = \left(
\begin{array}{c|c}
G_{2_2} & . \\
\hline
. & G_{2_2}
\end{array}
\right)
&\qquad \qquad& 
V_{(0,0),(0,1)} = \left(
\begin{array}{c|c}
. & G_{1_4} \\
\hline
G_{1_1} & G_{1_1}+G_{1_4}
\end{array}
\right)
\end{array}
\label{vije5}
\end{equation}

\paragraph{Realization of $Oc(\mathcal{E}_5)$}
The algebra of quantum symmetries $Oc(\mathcal{E}_5)$ can be realized as
\begin{equation}
Oc(\mathcal{E}_5) = \mathcal{E}_5 \otimes_J \mathcal{E}_5 \qquad \qquad
\textrm{with } a \otimes_J b.c = a.b^* \otimes_J c \qquad \textrm{for } b \in J = \{1_i\} \;,
\end{equation} 
where $J$ is a subalgebra characterized by modular properties (see \cite{Coq_Gil-Tmod,Gil-tesis}). 
The algebra $Oc(\mathcal{E}_5)$ has dimension $12 \times 2 = 24$, and a basis is given by elements $a \otimes_J 1_0$ and $a \otimes_J 2_0$. The identifications in $Oc(\mathcal{E}_5)$ are given by: 
\begin{equation}
\begin{array}{rcl} 
1_i \otimes_J 1_j &=& 1_{i+j^*} \otimes_J 1_0    \\ 
2_i \otimes_J 1_j &=& 2_{i+j^*} \otimes_J 1_0   \\ 
1_i \otimes_J 2_j  =  1_i \otimes_J 1_j . 2_0 &=& 1_{i+j^*} \otimes_J 2_0 \\ 
2_i \otimes_J 2_j  =  2_i \otimes_J 1_j . 2_0 &=& 2_{i+j^*} \otimes_J 2_0  
\end{array}
\label{ide5} 
\end{equation}
The chiral conjugation is defined by $(a \otimes_J b)^C = b \otimes_J a$. The left chiral generator is 
$2_1 \otimes_J 1_0$ and the right chiral generator is $1_0 \otimes_J 2_1 =  1_5 \otimes_J 2_0$. 
Multiplication in $Oc(\mathcal{E}_5)$ is defined from the multiplication (\ref{mult_E5}) of $\mathcal{E}_5$ together with the identifications (\ref{ide5}), and is encoded by the quantum symmetries matrices $O_x$. We get:
\begin{equation}
O_{x=a\otimes_J 1_0} = 
\left(
\begin{array}{rcl}
G_a & . \\
.   & G_a
\end{array}
\right)
\qquad \qquad \qquad
O_{x=a\otimes_J 2_0} = 
\left(
\begin{array}{rcl}
. & G_a \\
G_a.G_{1_3} & G_a (\munite + G_{1_3})
\end{array}
\right)
\label{oxe5}
\end{equation}
The vector space of $\mathcal{E}_5$ vertices is also a module under the action of vertices of $Oc(\mathcal{E}_5)$ defined by $(a\otimes_J 1_0) . b = a.b$ and $(a\otimes_J 2_0) . b = a.2_0.b$. The dual annular matrices $S_x$ are given by
$S_{x=a\otimes_J 1_0} = G_a$ and $S_{x=a\otimes_J 2_0}=G_{2_0}.G_a$. We check the dimensional rules $\dim(\mathcal{B}(\mathcal{E}_5)) = \sum_{\lambda} d_{\lambda}^2 = \sum_x d_x^2 = 29\,376$.


\subsection{The $\mathcal{E}_5^*$ case of $\widehat{su}(3)$}
We start now with the following  $\widehat{su}(3)_{5}$ modular invariant partition function:
\begin{eqnarray}
\nonumber \mathcal{Z} &=&  |\chi_{(0,0)}^5 + \chi_{(2,2)}^5|^2 + |\chi_{(3,0)}^5 + \chi_{(0,3)}^5|^2 +
[(\chi_{(0,2)}^5 + \chi_{(3,2)}^5).(\ov{\chi_{(2,0)}^5} + \ov{\chi_{(2,3)}^5}) + \textrm{h.c.}] \\
\label{modinve5c} && + (\chi_{(2,1)}^5 + \chi_{(0,5)}^5).(\ov{\chi_{(1,2)}^5} + \ov{\chi_{(5,0)}^5}) + \textrm{h.c.}]   \;,
\end{eqnarray}
and compute the modular matrix\footnote{Same remark as in the last footnote.} $\mathcal{M}$. The fusion matrices $N_{\lambda}$ are the same as in the previous case. 

\paragraph{Determination of toric matrices and double fusion matrices}
We have $d_O = Tr(\mathcal{M}\mathcal{M}^{\dagger}) = 24$. The matrices $K_{\lambda\mu}=N_{\lambda}\,\mathcal{M}\,N_{\mu}^{tr}$ span a vector space of dimension $r=d_O=24$.
The discussion is the same as in the previous case.  
\begin{itemize}
\item For squared norm 1 we have 21 linearly independent matrices $K_{\lambda\mu}$ defining 21 toric matrices $W_{z0}$.  
\item There are 45 linearly independent matrices $K_{\lambda\mu}$ of squared norm 2 and the last three toric matrices $W_{z0}$ can be obtained.  
\end{itemize}
Once the toric matrices have been determined, the double fusion matrices are obtained straightforwardly. For the fundamental ones we get:
\begin{equation}
\begin{array}{lcl}
V_{(1,0),(0,0)} = \left(
\begin{array}{c|c}
G_{2_1} & . \\
\hline
. & G_{2_1}
\end{array}
\right) 
&\qquad \qquad& 
V_{(0,0),(1,0)} = \left(
\begin{array}{c|c}
. & G_{1_1} \\
\hline
G_{1_4} & G_{1_1}+G_{1_4}
\end{array}
\right) 
 \\
{} & {} \\
V_{(0,1),(0,0)} = \left(
\begin{array}{c|c}
G_{2_2} & . \\
\hline
. & G_{2_2}
\end{array}
\right)
&\qquad \qquad& 
V_{(0,0),(0,1)} = \left(
\begin{array}{c|c}
. & G_{1_2} \\
\hline
G_{1_5} & G_{1_2}+G_{1_5}
\end{array}
\right)
\end{array}
\label{vije5conj}
\end{equation}

\begin{figure}[hhh] 
\begin{center} 
\unitlength 0.25mm 
 
\begin{picture}(400,410) 
 
\put(20,20){\begin{picture}(360,370)

\qbezier[10](0,325)(15,340)(0,350)
\qbezier[10](0,325)(-15,340)(0,350)

\qbezier[10](0,235)(15,250)(0,260)
\qbezier[10](0,235)(-15,250)(0,260)

\qbezier[10](90,190)(105,205)(90,215)
\qbezier[10](90,190)(75,205)(90,215)

\qbezier[10](90,370)(105,385)(90,395)
\qbezier[10](90,370)(75,385)(90,395)

\qbezier[10](180,325)(195,340)(180,350)
\qbezier[10](180,325)(165,340)(180,350)

\qbezier[10](180,235)(195,250)(180,260)
\qbezier[10](180,235)(165,250)(180,260)

\qbezier[10](240,135)(255,150)(240,160)
\qbezier[10](240,135)(225,150)(240,160)

\qbezier[10](300,135)(315,150)(300,160)
\qbezier[10](300,135)(285,150)(300,160)

\qbezier[10](210,90)(225,105)(210,115)
\qbezier[10](210,90)(195,105)(210,115)

\qbezier[10](330,90)(345,105)(330,115)
\qbezier[10](330,90)(315,105)(330,115)

\qbezier[10](240,45)(255,30)(240,20)
\qbezier[10](240,45)(225,30)(240,20)

\qbezier[10](300,45)(315,30)(300,20)
\qbezier[10](300,45)(285,30)(300,20)

\qbezier[80](60,235)(60,80)(180,135)
\qbezier[80](30,280)(30,90)(270,180)
\qbezier[80](120,235)(100,80)(180,45)

\qbezier[130](60,325)(250,470)(360,135)
\qbezier[140](120,325)(260,450)(360,45)
\qbezier[150](150,280)(700,80)(270,0)


\put(0,190){\begin{picture}(180,180)

\put(0,45){\begin{picture}(60,45) 
\put(0,0){\color{green} \circle*{7}} 
\put(60,0){\color{blue} \circle*{7}}
\put(30,45){\color{red} \circle*{7}} 
\put(0,0){\vector(1,0){32.5}} 
\put(30,0){\line(1,0){30}} 
\put(60,0){\vector(-2,3){16.5}} 
\put(45,22.5){\line(-2,3){15}} 
\put(30,45){\vector(-2,-3){16.5}} 
\put(15,22.5){\line(-2,-3){15}} 
\end{picture}} 
 
\put(120,45){\begin{picture}(60,45) 
\put(0,0){\color{green} \circle*{7}} 
\put(60,0){\color{blue} \circle*{7}}
\put(30,45){\color{red} \circle*{7}} 
\put(0,0){\vector(1,0){32.5}} 
\put(30,0){\line(1,0){30}} 
\put(60,0){\vector(-2,3){16.5}} 
\put(45,22.5){\line(-2,3){15}} 
\put(30,45){\vector(-2,-3){16.5}} 
\put(15,22.5){\line(-2,-3){15}} 
\end{picture}} 
 
\put(60,135){\begin{picture}(60,45) 
\put(0,0){\color{green} \circle*{7}} 
\put(60,0){\color{blue} \circle*{7}}
\put(30,45){\color{red} \circle*{7}} 
\put(0,0){\vector(1,0){32.5}} 
\put(30,0){\line(1,0){30}} 
\put(60,0){\vector(-2,3){16.5}} 
\put(45,22.5){\line(-2,3){15}} 
\put(30,45){\vector(-2,-3){16.5}} 
\put(15,22.5){\line(-2,-3){15}} 
\end{picture}}

\put(0,90){\begin{picture}(60,45) 
\put(0,45){\color{blue} \circle*{7}}
\put(60,45){\vector(-1,0){32.5}} 
\put(30,45){\line(-1,0){30}} 
\put(30,0){\vector(2,3){16.5}} 
\put(45,22.5){\line(2,3){15}} 
\put(0,45){\vector(2,-3){16.5}} 
\put(15,22.5){\line(2,-3){15}} 
\end{picture}} 
 
\put(120,90){\begin{picture}(60,45) 
\put(60,45){\color{green} \circle*{7}} 
\put(60,45){\vector(-1,0){32.5}} 
\put(30,45){\line(-1,0){30}} 
\put(30,0){\vector(2,3){16.5}} 
\put(45,22.5){\line(2,3){15}} 
\put(0,45){\vector(2,-3){16.5}} 
\put(15,22.5){\line(2,-3){15}} 
\end{picture}} 
 
\put(60,0){\begin{picture}(60,45) 
\put(30,0){\color{red} \circle*{7}} 
\put(60,45){\vector(-1,0){32.5}} 
\put(30,45){\line(-1,0){30}} 
\put(30,0){\vector(2,3){16.5}} 
\put(45,22.5){\line(2,3){15}} 
\put(0,45){\vector(2,-3){16.5}} 
\put(15,22.5){\line(2,-3){15}} 
\end{picture}} 
 
\put(60,135){\vector(0,-1){47.5}} 
\put(60,90){\line(0,-1){45}} 
\put(60,45){\vector(2,1){47.2}} 
\put(105,67.5){\line(2,1){45}} 
\put(150,90){\vector(-2,1){47.2}} 
\put(105,112.5){\line(-2,1){45}} 
 
\put(120,45){\vector(0,1){47.5}} 
\put(120,90){\line(0,1){45}} 
\put(120,135){\vector(-2,-1){47.2}} 
\put(75,112.5){\line(-2,-1){45}} 
\put(30,90){\vector(2,-1){47.2}} 
\put(75,67.5){\line(2,-1){45}}

\put(-30,47){\makebox(0,0){\scriptsize{$1_0\otimes 1_0$}}} 
\put(210,45){\makebox(0,0){\scriptsize{$1_4\otimes 1_0$}}} 
\put(-30,135){\makebox(0,0){\scriptsize{$1_1\otimes 1_0$}}} 
\put(210,135){\makebox(0,0){\scriptsize{$1_3\otimes 1_0$}}} 
\put(90,-12){\makebox(0,0){\scriptsize{$1_5\otimes 1_0$}}} 
\put(60,180){\makebox(0,0){\scriptsize{$1_2\otimes 1_0$}}} 

\put(45,35){\makebox(0,0){\scriptsize{$2_1\otimes 1_0$}}}  
\put(137,35){\makebox(0,0){\scriptsize{$2_0\otimes 1_0$}}} 
\put(43,148){\makebox(0,0){\scriptsize{$2_3\otimes 1_0$}}} 
\put(136,148){\makebox(0,0){\scriptsize{$2_4\otimes 1_0$}}} 
\put(5,92){\makebox(0,0){\scriptsize{$2_2\otimes 1_0$}}} 
\put(177,92){\makebox(0,0){\scriptsize{$2_5\otimes 1_0$}}} 

\end{picture}}


\put(180,0){\begin{picture}(180,180)

\put(0,45){\begin{picture}(60,45) 
\put(0,0){\color{green} \circle*{7}}
\put(60,0){\color{blue} \circle*{7}}
\put(30,45){\color{red} \circle*{7}} 
\put(0,0){\vector(1,0){32.5}} 
\put(30,0){\line(1,0){30}} 
\put(60,0){\vector(-2,3){16.5}} 
\put(45,22.5){\line(-2,3){15}} 
\put(30,45){\vector(-2,-3){16.5}} 
\put(15,22.5){\line(-2,-3){15}} 
\end{picture}} 
 
\put(120,45){\begin{picture}(60,45) 
\put(0,0){\color{green} \circle*{7}}
\put(60,0){\color{blue} \circle*{7}}
\put(30,45){\color{red} \circle*{7}} 
\put(0,0){\vector(1,0){32.5}} 
\put(30,0){\line(1,0){30}} 
\put(60,0){\vector(-2,3){16.5}} 
\put(45,22.5){\line(-2,3){15}} 
\put(30,45){\vector(-2,-3){16.5}} 
\put(15,22.5){\line(-2,-3){15}} 
\end{picture}} 
 
\put(60,135){\begin{picture}(60,45) 
\put(0,0){\color{green} \circle*{7}}
\put(60,0){\color{blue} \circle*{7}}
\put(30,45){\color{red} \circle*{7}} 
\put(0,0){\vector(1,0){32.5}} 
\put(30,0){\line(1,0){30}} 
\put(60,0){\vector(-2,3){16.5}} 
\put(45,22.5){\line(-2,3){15}} 
\put(30,45){\vector(-2,-3){16.5}} 
\put(15,22.5){\line(-2,-3){15}} 
\end{picture}}

\put(0,90){\begin{picture}(60,45) 
\put(0,45){\color{blue} \circle*{7}}
\put(60,45){\vector(-1,0){32.5}} 
\put(30,45){\line(-1,0){30}} 
\put(30,0){\vector(2,3){16.5}} 
\put(45,22.5){\line(2,3){15}} 
\put(0,45){\vector(2,-3){16.5}} 
\put(15,22.5){\line(2,-3){15}} 
\end{picture}} 
 
\put(120,90){\begin{picture}(60,45) 
\put(60,45){\color{green} \circle*{7}}
\put(60,45){\vector(-1,0){32.5}} 
\put(30,45){\line(-1,0){30}} 
\put(30,0){\vector(2,3){16.5}} 
\put(45,22.5){\line(2,3){15}} 
\put(0,45){\vector(2,-3){16.5}} 
\put(15,22.5){\line(2,-3){15}} 
\end{picture}} 
 
\put(60,0){\begin{picture}(60,45) 
\put(30,0){\color{red} \circle*{7}} 
\put(60,45){\vector(-1,0){32.5}} 
\put(30,45){\line(-1,0){30}} 
\put(30,0){\vector(2,3){16.5}} 
\put(45,22.5){\line(2,3){15}} 
\put(0,45){\vector(2,-3){16.5}} 
\put(15,22.5){\line(2,-3){15}} 
\end{picture}} 
 
\put(60,135){\vector(0,-1){47.5}} 
\put(60,90){\line(0,-1){45}} 
\put(60,45){\vector(2,1){47.2}} 
\put(105,67.5){\line(2,1){45}} 
\put(150,90){\vector(-2,1){47.2}} 
\put(105,112.5){\line(-2,1){45}} 
 
\put(120,45){\vector(0,1){47.5}} 
\put(120,90){\line(0,1){45}} 
\put(120,135){\vector(-2,-1){47.2}} 
\put(75,112.5){\line(-2,-1){45}} 
\put(30,90){\vector(2,-1){47.2}} 
\put(75,67.5){\line(2,-1){45}}

\put(-30,47){\makebox(0,0){\scriptsize{$1_0\otimes 2_0$}}} 
\put(210,45){\makebox(0,0){\scriptsize{$1_4\otimes 2_0$}}} 
\put(-30,135){\makebox(0,0){\scriptsize{$1_1\otimes 2_0$}}} 
\put(210,135){\makebox(0,0){\scriptsize{$1_3\otimes 2_0$}}} 
\put(90,-12){\makebox(0,0){\scriptsize{$1_5\otimes 2_0$}}} 
\put(90,195){\makebox(0,0){\scriptsize{$1_2\otimes 2_0$}}} 

\put(45,35){\makebox(0,0){\scriptsize{$2_1\otimes 2_0$}}}  
\put(137,35){\makebox(0,0){\scriptsize{$2_0\otimes 2_0$}}} 
\put(43,148){\makebox(0,0){\scriptsize{$2_3\otimes 2_0$}}} 
\put(136,148){\makebox(0,0){\scriptsize{$2_4\otimes 2_0$}}} 
\put(5,92){\makebox(0,0){\scriptsize{$2_2\otimes 2_0$}}} 
\put(177,92){\makebox(0,0){\scriptsize{$2_5\otimes 2_0$}}}

\end{picture}} 

\end{picture}}

\end{picture} 
\end{center} 
\caption{Ocneanu graph $Oc({\cal E}_5^*)$. The two left chiral generators are $2_1 \otimes 1_0$ and $2_2 \otimes 1_0$, the two right chiral generators are $1_1 \otimes 2_0$ and $1_2 \otimes 2_0$.} 
\label{OcE5conj} 
\end{figure}
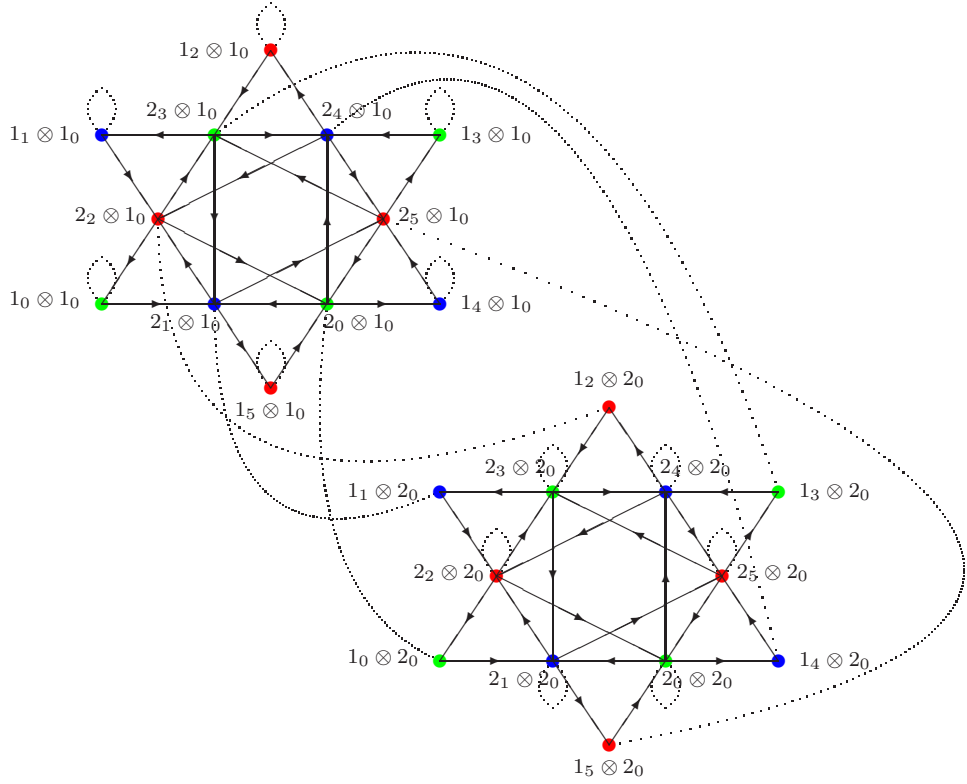

\paragraph{The Ocneanu graph of quantum symmetries}
We display in figure \ref{OcE5conj} the graph corresponding to the matrix $V_{(1,0),(0,0)}$ associated with the vertex labelled by $2_1\otimes1_0$. $V_{(0,0),(1,0)}$ is associated with the vertex $1_1 \otimes 2_0$. The algebra of quantum symmetries can be realized as
\begin{equation}
Oc(\mathcal{E}_5^*) = \mathcal{E}_5 \otimes_J \mathcal{E}_5 \qquad \qquad
\textrm{with } a \otimes_J b.c = a.b \otimes_J c \qquad \textrm{for } b \in J = \{1_i\} \;.
\end{equation}  
The algebra $Oc(\mathcal{E}_5^*)$ has also dimension $12 \times 2 = 24$ and a basis is given by elements $a \otimes_J 1_0$ and $a \otimes_J 2_0$. The identifications in $Oc(\mathcal{E}_5^*)$ are given by (different from those of $Oc(\mathcal{E}_5)$) 
\begin{equation}
\begin{array}{rcccl} 
1_i \otimes_J 1_j &=& 1_{i+j} \otimes_J 1_0    \\ 
2_i \otimes_J 1_j &=& 2_{i+j} \otimes_J 1_0   \\ 
1_i \otimes_J 2_j  =  1_i \otimes_J 1_j . 2_0 &=& 1_{i+j} \otimes_J 2_0  \\ 
2_i \otimes_J 2_j  =  2_i \otimes_J 1_j . 2_0 &=& 2_{i+j} \otimes_J 2_0 
\end{array}
\label{ide5conj} 
\end{equation}
The left chiral generator is $2_1 \otimes_J 1_0$ and the right chiral generator is 
$1_0 \otimes_J 2_1 = 1_1 \otimes_J 2_0$. 
The algebra $Oc(\mathcal{E}_5^*)$ is isomorphic to $Oc(\mathcal{E}_5)$, the quantum symmetry matrices $O_x$ are still given by (\ref{oxe5}). The difference is in the chiral conjugacy.

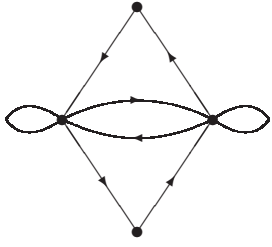
\begin{figure}[hhh] 
\begin{center} 
\unitlength 0.5mm 

\begin{picture}(40,60)(10,0)
\put(0,30){\circle*{3}} 
\put(40,30){\circle*{3}} 
\put(20,60){\circle*{3}}
\put(20,0){\circle*{3}}
 
\put(21,35){\vector(1,0){0}}
\put(18.5,25){\vector(-1,0){0}}

\put(28,48){\vector(-2,3){0}}
\put(10,45){\vector(-2,-3){0}}

\put(12,12){\vector(2,-3){0}}
\put(30,15){\vector(2,3){0}}

\qbezier(0,30)(20,40)(40,30)
\qbezier(0,30)(20,20)(40,30)

\qbezier(0,30)(-10,37)(-15,30)
\qbezier(0,30)(-10,23)(-15,30)
\qbezier(40,30)(50,37)(55,30)
\qbezier(40,30)(50,23)(55,30)

\put(20,60){\line(-2,-3){20}}
\put(20,60){\line(2,-3){20}} 
\put(20,0){\line(-2,3){20}}
\put(20,0){\line(2,3){20}}

\end{picture}
\end{center} 
\caption{The ${\cal E}_5^* = {\cal E}_5/3$ generalized Dynkin diagram.} 
\label{gre5conj} 
\end{figure}

\paragraph{The generalized Dynkin diagram $\mathcal{E}_5^*=\mathcal{E}_5/3$}
The graph associated to the initial modular invariant (\ref{modinve5c}) is a module graph for the Ocneanu graph displayed on figure \ref{OcE5conj}. It must therefore be a module graph of the $\mathcal{E}_5$ graph itself: it is obtained as the $Z_3$-orbifold graph of $\mathcal{E}_5$ (see \cite{Gil-maroc}). We write this module property $a \, \tilde{b} = 
\sum_{\tilde{c}} (F^{\mathcal{E}}_a)_{\tilde{b}\tilde{c}} \, \tilde{c}$, for $a \in \mathcal{E}_5$ and $\tilde{b},\tilde{c} \in \mathcal{E}_5/3$, encoded by the 12 matrices $F^{\mathcal{E}}_a$. From the associative property $(a . b) . \tilde{c} = a.(b.\tilde{c})$, these matrices must satisfy the same commutation relations (\ref{mult_E5}) as the graph algebra of $\mathcal{E}_5$, and can be recursively calculated from $F^{\mathcal{E}}_{2_1}$, which is the adjacency matrix of the $\mathcal{E}_5/3$ graph displayed on figure \ref{gre5conj}.  
The $\mathcal{E}_5/3$ graph is also a module over the algebra of quantum symmetries, the action being defined by $(a\otimes_J 1_0) . \tilde{b} = a. \tilde{b}$ and $(a\otimes_J 2_0) . b = a.2_0.\tilde{b}$. The dual annular matrices $S_x$ are therefore given by
$S_{x=a\otimes_J 1_0} = F^{\mathcal{E}}_a$ and $S_{x=a\otimes_J 2_0}=F^{\mathcal{E}}_{2_0}.F^{\mathcal{E}}_a$. We check the dimensional rules $\dim(\mathcal{B}(\mathcal{E}_5^*)) = \sum_{\lambda} d_{\lambda}^2 = \sum_x d_x^2 = 3\,264$.

So both graphs $G=\mathcal{E}_5$ and $\mathcal{E}_5/3$ have the same (isomorphic) algebra $Oc(G)$ of quantum symmetries, but its realization in terms of tensor square of $\mathcal{E}_5$ is different in the two cases, as well as the chiral conjugation, and, of course, its module action on $\mathcal{E}_5$ or on $\mathcal{E}_5/3$. 


\subsection{The $\mathcal{E}_9$ case of $su(3)$}
We start with the following $\widehat{su}(3)_{9}$ modular invariant partition function:
\begin{equation}
\mathcal{Z} =  |\chi_{0,0}^9 + \chi_{0,9}^9 + \chi_{9,0}^9 + \chi_{1,4}^9 +
\chi_{4,1}^9 + \chi_{4,4}^9|^2 + 2 \, |\chi_{2,2}^9 + \chi_{2,5}^9 + \chi_{5,2}^9|^2 \;,
\label{modinve9}
\end{equation}
where $\chi_{\lambda}^9$'s are the characters of $\widehat{su}(3)_{9}$, labelled by $\lambda=(\lambda_1,\lambda_2)$ with $0\leq \lambda_1,\lambda_2\leq 9$, $\lambda_1+\lambda_2\le 9$. Notice that this modular invariant can be obtained from the conformal embedding of affine algebras $\widehat{su}(3)_9 \subset (\widehat{E}_6)_1$.  The modular invariant matrix is recovered from   $\mathcal{Z}=\sum_{\lambda}\chi_{\lambda}\mathcal{M}_{\lambda\mu}\overline{\chi}_{\mu}$.
The number of irreps is $d_{\mathcal{A}}= 55$. The fusion matrix $N_{(1,0)}$ is the adjacency matrix of the $\mathcal{A}_9$ graph, the truncated Weyl alcove of SU(3) irreps at level 9. The other fusion matrices are determined by the recurrence relation (\ref{su3recrel}). 

\paragraph{Determination of toric matrices $W_{z0}$} We have $d_O = Tr(\mathcal{M}\mathcal{M}^{\dagger}) = 72$ and therefore an Ocneanu algebra with 72 generators $z$ and also 72 toric matrices $W_{z0}$. However these toric matrices span a vector space of dimension $r=45<72$, i.e. they are not all linearly independent. For each matrix  $K_{\lambda\mu} = N_{\lambda}\mathcal{M}N_{\mu}^{tr}$ we consider its ``squared norm'' defined by $||K_{\lambda\mu}||^2 = (K_{\lambda\mu})_{\lambda^*\mu^*}$:
\begin{itemize}
\item[$\bullet$] There are 27 matrices $K_{\lambda\mu}$ with squared norm 1, each one defines a toric matrix $W_{z0}$. 

\item[$\bullet$] There are 12 linearly independent matrices $K_{\lambda\mu}$ with squared norm 2, but each one is equal to the sum of two already determined matrices. We don't find any new toric matrix in this family. 

\item[$\bullet$] There are 21 linearly independent matrices $K_{\lambda\mu}$ of squared norm 3, none of them being equal to the sum of three already obtained matrices. Twelve amoung these 21 are equal to the sum of one determined matrix and a matrix having coefficients multiple of 2. A solution leading to squared norm 3 is to define a new toric matrix by dividing by 2 the matrix with coefficients multiple of 2, and adding them to the list with a multiplicity two. From these twelve we obtain actually only  eight different toric matrices (because some are obtained more than once), each one coming with multiplicity two.   
Nine of the 21 matrices have coefficients which are multiple of 3. We define nine new toric matrices by dividing these matrices by 3, each toric matrix obtained in that way appearing with multiplicity 3. At that stage, we have determined $27 + (2\times 8) + (3\times 9) = 70$ toric matrices.  

\item[$\bullet$] There are 24 linearly independent matrices $K_{\lambda\mu}$ with squared norm 4, but each one is equal to the sum of four already obtained matrices. We don't recover any new toric matrix. This is also the case for squared norm 5. 

\item[$\bullet$] There are 10 linearly independent matrices $K_{\lambda\mu}$ with squared norm 6. We discard those that can be written as a linear combination of already determined toric matrices, and pick up one of the others, for example $K_{ab}$. We build the list of matrices $K_{ab}-W_x$, for $W_x$ running into the set of already obtained toric matrices, searching for matrices with non-negative coefficients. With our choice, it is so that $K_{ab}$ is the sum of two times a toric matrix plus a new one which has matrix elements multiple of 2. Dividing the later by 2 and adding it to the list, with multiplicity 2, we get the last toric matrices. 
\end{itemize}
We have indeed therefore determined the $72$ toric matrices, 45 (=27+9+8+1) of them being linearly independent, but appearing with multiplicities (27 of multiplicity one, 9 (=8+1) of multiplicity two and 9 of multiplicity three). We can check the result by a direct substitution in the $55\times 55 = 3025$ matrix equations over non-negative integers (\ref{Kmse}).

\paragraph{Determination of $V_{(1,0),(0,0)}$ and $V_{(0,0),(1,0)}$}
We compute the set of matrices $K_{\lambda\mu}^x = N_{\lambda}W_{x0}N_{\mu}^{tr}$ for $\{\lambda\mu\}=\{(1,0),(0,0)\}$
and $\{(0,0),(1,0)\}$, and decompose them on the family (not a base) of toric matrices $W_{z0}$ using (\ref{Kmse}). Since the $W_{z0}$ are not linearly independent, the decomposition is not unique, and we introduce some undetermined coefficients. Imposing that they should be non-negative integers allows to fix some of them or to obtain relations between them. More constraints come from the fact that we have $V_{(0,0),(1,0)}=C.V_{(0,0),(1,0)}.C^{-1}$, where $C$ is the chiral operator.
Notice that $C$ itself is deduced from the previous relation even if $V_{(0,0),(1,0)}$ and $V_{(0,0),(1,0)}$ still contain free parameters, by using the fact that it is a permutation matrix. 
Choosing an appropriate order on the set of indices $z$, we obtain the following structure for $V_{(1,0),(0,0)}$:
\begin{equation}
V_{(1,0),(0,0)} = \left(
\begin{array}{cccccc}
Ad(\mathcal{E}_9) & . & . & . & . & .\\
. & Ad(\mathcal{E}_9) & . & . & . & .\\
. & . & Ad(\mathcal{E}_9) & . & . & .\\
. & . & . & Ad(\mathcal{M}_9) & . & .\\
. & . & . & . & Ad(\mathcal{M}_9) & .\\
. & . & . & . & . & Ad(\mathcal{M}_9)\\
\end{array}
\right)
\label{V10}
\end{equation}
where  $Ad(\mathcal{E}_9)$ and $Ad(\mathcal{M}_9)$ are $12\times 12$ matrices (still containing some unknown coefficients).

\begin{figure}
 \psfrag{1}{\footnotesize$0_0$}
 \psfrag{2}{\footnotesize$1_0$}
  \psfrag{3}{\footnotesize$2_0$}
  \psfrag{4}{\footnotesize$3_0$}
  \psfrag{5}{\footnotesize$0_1$}
  \psfrag{6}{\footnotesize$1_1$}
  \psfrag{7}{\footnotesize$2_1$}
  \psfrag{8}{\footnotesize$2_1$}
  \psfrag{9}{\footnotesize$0_2$}
  \psfrag{10}{\footnotesize$1_2$}
  \psfrag{11}{\footnotesize$2_2$}
  \psfrag{12}{\footnotesize$3_2$}
  \psfrag{1m}{\footnotesize$\tilde0_0$}
  \psfrag{2m}{\footnotesize$\tilde3_0$}
  \psfrag{3m}{\footnotesize$\tilde3'_0$}
  \psfrag{4m}{\footnotesize$\tilde3''_0$}
  \psfrag{5m}{\footnotesize$\tilde0_1$}
  \psfrag{6m}{\footnotesize$\tilde3_1$}
  \psfrag{7m}{\footnotesize$\tilde3'_1$}
  \psfrag{8m}{\footnotesize$\tilde3''_1$}
  \psfrag{9m}{\footnotesize$\tilde0_2$}
  \psfrag{10m}{\footnotesize$\tilde3_2$}
  \psfrag{11m}{\footnotesize$\tilde3'_2$}
  \psfrag{12m}{\footnotesize$\tilde3''_2$}
\psfrag{E9}{\Large$\mathcal{E}_9$}
\psfrag{M9}{\Large$\mathcal{M}_9$}
\centerline{\includegraphics[width=150mm,height=80mm]{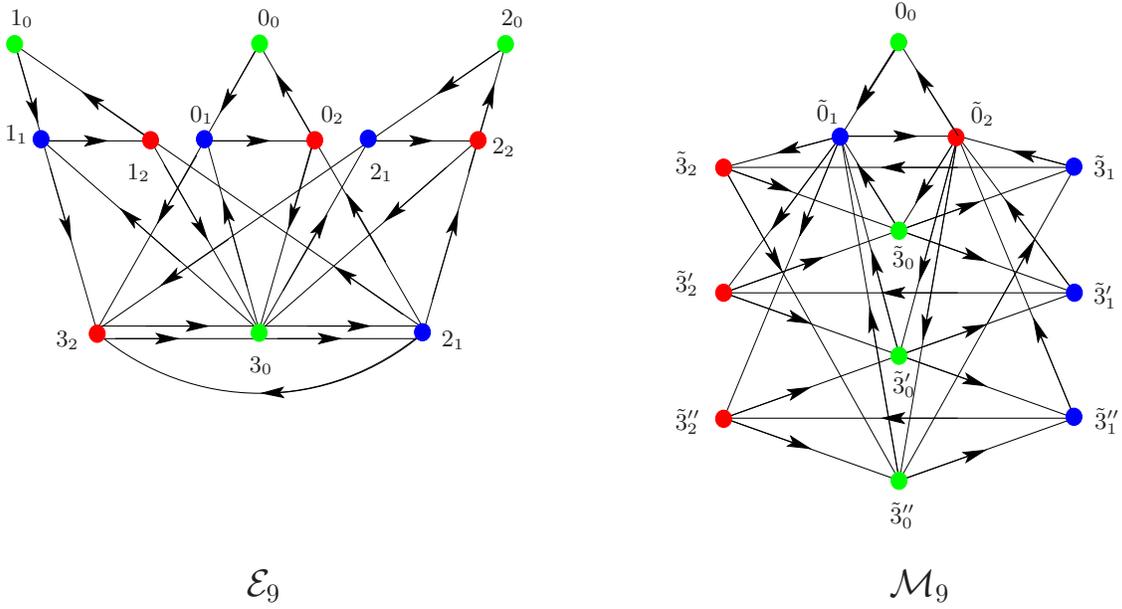}}
\caption{The graphs $\mathcal{E}_9$ and $\mathcal{M}_9$}
\label{grE9M9} 
\end{figure}

\paragraph{The generalized Dynkin diagram $\mathcal{E}_9$}
The $Ad(\mathcal{E}_9)$ matrix is the adjacency matrix of the graph $\mathcal{E}_9$ displayed on the l.h.s. of figure 
\ref{grE9M9}. It possesses a $\mathbb{Z}_3$-symmetry corresponding to the permutation of the three ``wings'' formed by vertices $0_i$, $1_i$ and $2_i$. The undetermined coefficients of the adjacency matrix reflect this symmetry; they are simply fixed once an ordering has been chosen for the vertices (something similar happens for the $D_{even}$ series of the $su(2)$ family). 

The vector space of the $\mathcal{E}_9$ graph is a module over the left-right action of the graph algebra of the $\mathcal{A}_9$ graph, encoded by the annular matrices $F^{\mathcal{E}}_{\lambda}$
\begin{equation}
\mathcal{A}_9 \times \mathcal{E}_9 \rightarrow \mathcal{E}_9 : \quad \lambda \cdot a = a \cdot \lambda =
 \sum_{b} (F^{\mathcal{E}}_{\lambda})_{ab}\;b \qquad \qquad \lambda \in \mathcal{A}_9\;, \quad a,b \in \mathcal{E}_9 \;.
\end{equation}
The $F^{\mathcal{E}}_{\lambda}$ matrices give a representation of dimension 12 of the fusion algebra and are determined from the recursion relation (\ref{su3recrel}) with $F^{\mathcal{E}}_{(0,0)}=\munite_{12\times 12}$, $F^{\mathcal{E}}_{(1,0)} = Ad(\mathcal{E}_9)$.
We notice that fundamental matrices (for instance $F_{(1,0)}$) contain, in this case, elements bigger than 1, however, the ``rigidity\footnote{We call it that way  because of its relation with the theory of rigid categories (see for instance \cite{Ost}).}  condition'' $(F_\lambda)_{ab}= (F_{\lambda^*})_{ba}$ holds, so that this example is indeed an higher analogue of the ADE graphs, not an higher analogue of the non simply laced cases. 
Triality and conjugation compatible with the action of $\mathcal{A}_9$ can be defined on the $\mathcal{E}_9$ graph. 
Triality is denoted by the index $i \in \{0,1,2\}$ in the set of vertices $0_i,1_i,2_i$. The conjugation corresponds to the vertical axis going through vertices $0_0$ and $3_0$: $0_0^*=0_0, 1_0^*=2_0, 3_0^*=3_0$, 
$0_1^*=0_2,1_1^*=2_2,1_2^*=2_1,3_1^*=3_2$. 
The $\mathbb{Z}_3$-symmetry action on vertices of $\mathcal{E}_9$ is denoted $\rho_3$. The axis formed by vertices $3_i$ is invariant under $\rho_3$ and the symmetry permutes the three wings $\rho_3(0_0)=1_0$,
$\rho_3(1_0)=2_0$, $\rho_3(2_0)=0_0$; $\rho_3(0_1)=1_1$, $\rho_3(1_1)=2_1$, $\rho_3(2_1)=0_1$; $\rho_3(0_2)=1_2$, $\rho_3(1_2)=2_2$, $\rho_3(2_2)=0_2$. Once we have fixed the origin of the graph (the vertex $0_0$), the graph still possesses a $\mathbb{Z}_2$-symmetry corresponding to the permutation of the two remaining wings, formed by vertices $1_i$ and $2_i$. We denote $\rho_2$ this operation: $\rho_2(1_i)=2_i$ and $\rho_2^2=\munite$.

The $\mathcal{E}_9$ graph has also self-fusion: the vector space spanned by its vertices has an associative algebra structure, with non-negative structure constants, compatible with the action of $\mathcal{A}_9$. $0_0$ is the unity and the two conjugated generators are $0_1$ and $0_2$. The graph itself is also the Cayley graph of  multiplication by $0_1$.  Due to the symmetry of the wings of the graph, the knowledge of the multiplication by generators $0_1$ and $0_2$ is not sufficient to reconstruct the whole multiplication table; we have to impose structure coefficients to be non-negative integers in order to determine a unique solution (see \cite{Coq_Gil-Tmod, Gil-tesis}). The whole multiplication table is encoded in the graph algebra matrices $G_a$, for $a \in \mathcal{E}_9$.
We give the expression for $G_{1_0}$ and $G_{2_0}$, the other matrices are computed by $G_{0_0}=\munite$, 
$G_{0_1}=G_{0_2}^{tr}=Ad(\mathcal{E}_9)$, 
$G_{3_0}=G_{0_1}\, G_{0_2}-G_{0_0}$, 
$G_{3_2}=G^{tr}_{3_1}=G_{0_1}\, G_{0_1}-G_{0_2}$,
$G_{1_1}=G^{tr}_{2_2}=G_{0_1}\, G_{1_0}$, 
$G_{1_2}=G^{tr}_{2_1}=G_{0_2}\, G_{1_0}$. 
In the ordered basis $(0_0,1_0,2_0,3_0;  
0_1,1_1,2_1,3_1; 0_2,1_2,2_2,3_2)$, $G_{1_0}$ and $G_{2_0}$ are given by 
\begin{equation}
G_{{1}_{0}}  =  G_{{2}_{0}}^{tr} = 
\left( \begin{array}{cccccccccccc} 
. & 1 & . & . & . & . & . & . & . & . & . & .  \\ 
. & . & 1 & . & . & . & . & . & . & . & . & .  \\ 
1 & . & . & . & . & . & . & . & . & . & . & .  \\ 
. & . & . & 1 & . & . & . & . & . & . & . & .  \\ 
. & . & . & . & . & 1 & . & . & . & . & . & .  \\ 
. & . & . & . & . & . & 1 & . & . & . & . & .  \\ 
. & . & . & . & 1 & . & . & . & . & . & . & .  \\ 
. & . & . & . & . & . & . & 1 & . & . & . & .  \\ 
. & . & . & . & . & . & . & . & . & 1 & . & .  \\ 
. & . & . & . & . & . & . & . & . & . & 1 & .  \\ 
. & . & . & . & . & . & . & . & 1 & . & . & .  \\ 
. & . & . & . & . & . & . & . & . & . & . & 1  
\end{array} 
\right) 
\end{equation}
Notice that multiplication by $1_0$ corresponds to the $\mathbb{Z}_3$ operation: $1_0 . a = \rho_3(a)$. The matrix $G_{1_0}$ is the permutation matrix representing the action of the $\mathbb{Z}_3$ operator $\rho_3$:
$(G_{1_0})_{ab} = \delta_{b,\rho_3(a)}$. We have $(G_{1_0})^3=\munite$ and $(G_{1_0})^2=G_{2_0}$, so $G_{2_0}$ represents the operator $(\rho_3)^2$. 

Other aspects and properties of the $\mathcal{E}_9$ graph and of its algebra of quantum symmetries (semi-simple structure of the associated quantum groupo\"id, semi-simple structure of $Oc(\mathcal{E}_9)$ itself, quantum dimensions and quantum mass) are presented in \cite{Coq_Gil-Tmod,Gil-tesis,Coq-maroc}.

\paragraph{The generalized Dynkin diagram $\mathcal{M}_9$}
The matrix $Ad(\mathcal{M}_9)$ is a $12\times 12$ matrix with some unknown coefficients to be determined. Imposing this matrix to be the adjacency matrix of a graph such that the vector space spanned by its vertices is a module over the graph algebras of $\mathcal{A}_9$ {\sl and} of $\mathcal{E}_9$ leads to a unique solution. The graph is displayed on the r.h.s. of figure 
\ref{grE9M9} and corresponds to the $\mathbb{Z}_3$-orbifold graph of $\mathcal{E}_9$, denoted $\mathcal{M}_9 = \mathcal{E}_9/3$. 

The vector space spanned by vertices of the $\mathcal{M}_9$ graph is a module over the left-right action of the graph algebra of $\mathcal{A}_9$ encoded by the annular matrices $F^{\mathcal{M}}_{\lambda}$
\begin{equation}
\mathcal{A}_9 \times \mathcal{M}_9 \rightarrow \mathcal{M}_9 : \quad \lambda \cdot \tilde{a} =  \tilde{a} \cdot \lambda = 
 \sum_{\tilde{b}} (F^{\mathcal{M}}_{\lambda})_{\tilde{a}\tilde{b}}\;\tilde{b} \qquad \qquad \lambda \in \mathcal{A}_9\;, \quad \tilde{a},\tilde{b} \in \mathcal{M}_9 \;.
\end{equation}
The $F^{\mathcal{M}}_{\lambda}$ matrices give a representation of dimension 12 of the fusion algebra and can be determined from the  recursion relation (\ref{su3recrel}) with $F^{\mathcal{M}}_{(0,0)}=\munite_{12\times 12}$, $F^{\mathcal{M}}_{(1,0)} = Ad(\mathcal{M}_9)$. Triality and conjugation compatible with the action of $\mathcal{A}_9$ can be defined on the $\mathcal{M}_9$ graph. Triality is denoted by the index $i \in \{0,1,2\}$ in the set of vertices $\tilde{a}_i \in \mathcal{M}_9$. The conjugation corresponds to the vertical axis going through vertex $\tilde{0}_0$: $\tilde{0}_0^*=\tilde{0}_0, \tilde{0}_1^*=\tilde{0}_2, \tilde{3}_0^*=\tilde{3}_0$, 
${\tilde{3}_0'}{}^{*}=\tilde{3}_0'$, ${\tilde{3}_0''}{}^*=\tilde{3}_0'',\tilde{3}_1^*=\tilde{3}_2,{\tilde{3}_1'}{}^*=\tilde{3}_2', 
{\tilde{3}_1''}{}^*=\tilde{3}_2''$.

The vector space spanned by vertices of $\mathcal{M}_9$ is also a module under the action of the graph algebra of $\mathcal{E}_9$. Here we will distinguish between left and right action. 
The left action of $\mathcal{E}_9$ is encoded by a set of $12\times12$ matrices denoted $P^{\ell}_{\lambda}$
\begin{equation}
\mathcal{E}_9 \times \mathcal{M}_9 \rightarrow \mathcal{M}_9 : \quad a \cdot \tilde{b} = 
 \sum_{\tilde{c}} (P^{\ell}_{a})_{\tilde{b}\tilde{c}}\;\tilde{c} \qquad \qquad a \in \mathcal{E}_9\;, \quad 
\tilde{b},\tilde{c} \in \mathcal{M}_9 \;.
\end{equation}
The module property $(a\cdot b)\cdot \tilde{c}=a\cdot(b\cdot\tilde{c})$ imposes $P^{\ell}_{a}$ matrices
to form a representation of the graph algebra of $\mathcal{E}_9$; they satisfy  
$P^{\ell}_{a} \, P^{\ell}_{b} = \sum_c (G_{a})_{bc} P^{\ell}_{c}$.  We compute the set of matrices $P^{\ell}_{a}$ using the multiplicative structure of $\mathcal{E}_9$ from the previous relation. We give below the expression for $P^{\ell}_{1_0}$ and $P^{\ell}_{2_0}$, the other matrices being computed by $P^{\ell}_{0_0}=\munite$, 
$P^{\ell}_{0_1}=(P^{\ell}_{0_2})^{tr}=Ad(\mathcal{M}_9)$, 
$P^{\ell}_{3_0}=P^{\ell}_{0_1}  P^{\ell}_{0_2}-P^{\ell}_{0_0}$, 
$P^{\ell}_{3_2}=(P^{\ell}_{3_1})^{tr}=P^{\ell}_{0_1}\, P^{\ell}_{0_1}-P^{\ell}_{0_2}$,
$P^{\ell}_{1_1}=(P^{\ell}_{2_2})^{tr}=P^{\ell}_{0_1}\, P^{\ell}_{1_0}$, 
$P^{\ell}_{1_2}=(P^{\ell}_{2_1})^{tr}=P^{\ell}_{0_2}\, P^{\ell}_{1_0}$. 
In the ordered basis $(\tilde{0}_0,\tilde{3}_0,\tilde{3}_0',\tilde{3}_0'';  
\tilde{0}_1,\tilde{3}_1,\tilde{3}_1',\tilde{3}_1''; \tilde{0}_2,\tilde{3}_2,\tilde{3}_2',\tilde{3}_2'')$, $P^{\ell}_{1_0}$ and $P^{\ell}_{2_0}$ are given by
\begin{equation}
\begin{array}{ccc}
P^{\ell}_{1_0}=(P^{\ell}_{2_0})^{tr}=
\left(
\begin{array}{llllllllllll}
 1 & . & . & . & . & . & . & . & . & . & . & . \\
 . & . & 1 & . & . & . & . & . & . & . & . & . \\
 . & . & . & 1 & . & . & . & . & . & . & . & . \\
 . & 1 & . & . & . & . & . & . & . & . & . & . \\
 . & . & . & . & 1 & . & . & . & . & . & . & . \\
 . & . & . & . & . & . & 1 & . & . & . & . & . \\
 . & . & . & . & . & . & . & 1 & . & . & . & . \\
 . & . & . & . & . & 1 & . & . & . & . & . & . \\
 . & . & . & . & . & . & . & . & 1 & . & . & . \\
 . & . & . & . & . & . & . & . & . & . & 1 & . \\
 . & . & . & . & . & . & . & . & . & . & . & 1 \\
 . & . & . & . & . & . & . & . & . & 1 & . & .
\end{array}
\right)
\end{array}
\label{pEl}
\end{equation}
There is also an operator $\rho_3'$ acting on vertices of the $\mathcal{M}_9$ graph, inherited from the $\mathbb{Z}_3$ symmetry of the $\mathcal{E}_9$ graph through the orbifold procedure. It satisfies the following property:
\begin{equation}
\rho_3(a) \, \tilde{b} = a \, \rho_3'(\tilde{b})
\end{equation}
We have $1_0 \, a = \rho_3(a)$, so $\rho_3'(\tilde{a}) = 1_0 \, \tilde{a}$. 
It is defined by $\rho_3'(\tilde{0}_i)=\tilde{0}_i$,
$\rho_3'(\tilde{3}_i)=\tilde{3}_i'$, $\rho_3'(\tilde{3}_i')=\tilde{3}_i''$, $\rho_3'(\tilde{3}_i'')=\tilde{3}_i$, for $i=0,1,2$. The matrix $P^{\ell}_{1_0}$ is therefore the permutation matrix representing the action of the $\mathbb{Z}_3$ operator $\rho_3'$. We have $(P^{\ell}_{1_0})^3=\munite$ and $(P^{\ell}_{1_0})^2=P^{\ell}_{2_0}$, so $P^{\ell}_{2_0}$ represents the operator $(\rho_3')^2$.

\paragraph{The vector space $\mathcal{E}_9 \oplus \mathcal{M}_9$} 
We define the vector space $H = \mathcal{E}_9 \oplus \mathcal{M}_9$, and we want to define (this will be used later) an associative product on $H$ with the following structure:
$$
\begin{array}{c|cc}
\nearrow & \mathcal{E}_9 & \mathcal{M}_9 \\
\hline
\mathcal{E}_9 & \mathcal{E}_9 & \mathcal{M}_9 \\
\mathcal{M}_9 & \mathcal{M}_9 & \mathcal{E}_9 \\
\end{array}
$$
We define the following actions:
\begin{equation}
\begin{array}{rcl}
\mathcal{E}_9 \times \mathcal{E}_9 \rightarrow \mathcal{E}_9 &:& \displaystyle a \, b = \sum_c (G_a)_{bc} \, c \\
\mathcal{E}_9 \times \mathcal{M}_9 \rightarrow \mathcal{M}_9 &:& \displaystyle a \, \tilde{b} = \sum_{\tilde{c}} (P^{\ell}_a)_{\tilde{b}\tilde{c}} \, \tilde{c} \\
\mathcal{M}_9 \times \mathcal{E}_9 \rightarrow \mathcal{M}_9 &:& \displaystyle \tilde{b} \, a = \sum_{\tilde{c}} (P^{r}_a)_{\tilde{b}\tilde{c}} \, \tilde{c} \\
\mathcal{M}_9 \times \mathcal{M}_9 \rightarrow \mathcal{E}_9 &:& \displaystyle \tilde{a} \, \tilde{b} = \sum_c (H_{\tilde{a}})_{\tilde{b}c} \, c .
\end{array}
\label{multh}
\end{equation}
The associativity property on $H$ reads  
$a \,(b \, c) = (a \, b) \, c$ ; 
$a \,(b \, \tilde{c}) = (a \, b) \, \tilde{c}$ ;
$a \,(\tilde{b} \, c) = (a \, \tilde{b}) \, c$ ;
$\tilde{a} \,(b \, c) = (\tilde{a} \, b) \, c$ ;
$a \,(\tilde{b} \, \tilde{c}) = (a \, \tilde{b}) \, \tilde{c}$ ;
$\tilde{a} \,(b \, \tilde{c}) = (\tilde{a} \, b) \, \tilde{c}$ ;
$\tilde{a} \,(\tilde{b} \, c) = (\tilde{a} \, \tilde{b}) \, c$ ;
$\tilde{a} \,(\tilde{b} \, \tilde{c}) = (\tilde{a} \, \tilde{b}) \, \tilde{c}$, and induce a set of relations between matrices $G_a, P_a^{\ell},P_a^r$ and $H_{\tilde{a}}$. In order to satisfy them we found a unique solution for matrices $P_a^r$ and $H_{\tilde{a}}$. 
The right action of $\mathcal{E}_9$ on $\mathcal{M}_9$ encoded by the set of matrices $P^{r}_{a}$ is defined via the $\mathbb{Z}_2$ operator $\rho_2$:
\begin{equation}
\tilde{b} \cdot a = \rho_2(a) \cdot \tilde{b}
\end{equation} 
so that we have $P^r_a = P^{\ell}_{\rho_2(a)}$.
The coefficients of the $H_{\tilde{a}}$ matrices are given by:
\begin{equation}
(H_{\tilde{a}})_{\tilde{b}c} = (P^{\ell}_{\rho_2(c)})_{\tilde{a}^*\tilde{b}}= (P^{r}_{c})_{\tilde{a}^*\tilde{b}} \;.
\end{equation}

\begin{figure}

\psfrag{1}{\footnotesize$1_00_0$}
 \psfrag{2}{\footnotesize$0_00_0$}
 \psfrag{3}{\footnotesize$2_00_0$}
 \psfrag{4}{\footnotesize$1_10_0$}
 \psfrag{5}{\footnotesize$1_20_0$}
 \psfrag{6}{\footnotesize$0_10_0$}
 \psfrag{7}{\footnotesize$0_20_0$}
 \psfrag{8}{\footnotesize$2_10_0$}
 \psfrag{9}{\footnotesize$2_20_0$}
 \psfrag{10}{\footnotesize$3_20_0$}
 \psfrag{11}{\footnotesize$3_00_0$}
 \psfrag{12}{\footnotesize$3_10_0$}
 \psfrag{13}{\footnotesize$1_00_1$}
 \psfrag{14}{\footnotesize$0_00_1$}
 \psfrag{15}{\footnotesize$2_00_1$}
 \psfrag{16}{\footnotesize$1_10_1$}
 \psfrag{17}{\footnotesize$1_20_1$}
 \psfrag{18}{\footnotesize$0_10_1$}
 \psfrag{19}{\footnotesize$0_20_1$}
 \psfrag{20}{\footnotesize$2_10_1$}
 \psfrag{21}{\footnotesize$2_20_1$}
 \psfrag{22}{\footnotesize$3_20_1$}
 \psfrag{23}{\footnotesize$3_00_1$}
 \psfrag{24}{\footnotesize$3_10_1$}
 \psfrag{25}{\footnotesize$1_00_2$}
 \psfrag{26}{\footnotesize$0_00_2$}
 \psfrag{27}{\footnotesize$2_00_2$}
 \psfrag{28}{\footnotesize$1_10_2$}
 \psfrag{29}{\footnotesize$1_20_2$}
 \psfrag{30}{\footnotesize$0_10_2$}
 \psfrag{31}{\footnotesize$0_20_2$}
 \psfrag{32}{\footnotesize$2_10_2$}
 \psfrag{33}{\footnotesize$2_20_2$}
 \psfrag{34}{\footnotesize$3_20_2$}
 \psfrag{35}{\footnotesize$3_00_2$}
 \psfrag{36}{\footnotesize$3_10_2$}
 \psfrag{37}{\footnotesize$\tilde{0}_0\tilde{3}_2$}
 \psfrag{38}{\footnotesize$\tilde{0}_1\tilde{3}_2$}
 \psfrag{39}{\footnotesize$\tilde{0}_2\tilde{3}_2$}
 \psfrag{40}{\footnotesize$\tilde{3}_2\tilde{3}_2$}
 \psfrag{41}{\footnotesize$\tilde{3}_1\tilde{3}_2$}
 \psfrag{42}{\footnotesize$\tilde{3}'_2\tilde{3}_2$}
 \psfrag{43}{\footnotesize$\tilde{3}'_1\tilde{3}_2$}
 \psfrag{44}{\footnotesize$\tilde{3}''_2\tilde{3}_2$}
 \psfrag{45}{\footnotesize$\tilde{3}''_1\tilde{3}_2$}
 \psfrag{46}{\footnotesize$\tilde{3}_0\tilde{3}_2$}
 \psfrag{47}{\footnotesize$\tilde{3}'_0\tilde{3}_2$}
 \psfrag{48}{\footnotesize$\tilde{3}''_0\tilde{3}_2$}
\psfrag{49}{\footnotesize$\tilde{0}_0\tilde{3}_0$}
 \psfrag{50}{\footnotesize$\tilde{0}_1\tilde{3}_0$}
 \psfrag{51}{\footnotesize$\tilde{0}_2\tilde{3}_0$}
 \psfrag{52}{\footnotesize$\tilde{3}_2\tilde{3}_0$}
 \psfrag{53}{\footnotesize$\tilde{3}_1\tilde{3}_0$}
 \psfrag{54}{\footnotesize$\tilde{3}'_2\tilde{3}_0$}
 \psfrag{55}{\footnotesize$\tilde{3}'_1\tilde{3}_0$}
 \psfrag{56}{\footnotesize$\tilde{3}''_2\tilde{3}_0$}
 \psfrag{57}{\footnotesize$\tilde{3}''_1\tilde{3}_0$}
 \psfrag{58}{\footnotesize$\tilde{3}_0\tilde{3}_0$}
 \psfrag{59}{\footnotesize$\tilde{3}'_0\tilde{3}_0$}
 \psfrag{60}{\footnotesize$\tilde{3}''_0\tilde{3}_0$}
\psfrag{61}{\footnotesize$\tilde{0}_0\tilde{3}_1$}
 \psfrag{62}{\footnotesize$\tilde{0}_1\tilde{3}_1$}
 \psfrag{63}{\footnotesize$\tilde{0}_2\tilde{3}_1$}
 \psfrag{64}{\footnotesize$\tilde{3}_2\tilde{3}_1$}
 \psfrag{65}{\footnotesize$\tilde{3}_1\tilde{3}_1$}
 \psfrag{66}{\footnotesize$\tilde{3}'_2\tilde{3}_1$}
 \psfrag{67}{\footnotesize$\tilde{3}'_1\tilde{3}_1$}
 \psfrag{68}{\footnotesize$\tilde{3}''_2\tilde{3}_1$}
 \psfrag{69}{\footnotesize$\tilde{3}''_1\tilde{3}_1$}
 \psfrag{70}{\footnotesize$\tilde{3}_0\tilde{3}_1$}
 \psfrag{71}{\footnotesize$\tilde{3}'_0\tilde{3}_1$}
 \psfrag{72}{\footnotesize$\tilde{3}''_0\tilde{3}_1$}
 \psfrag{E9}{\Large$\mathcal{E}_9$}
 \psfrag{E9r}{\Large$\mathcal{E}^r_9$}
 \psfrag{E9l}{\Large$\mathcal{E}^l_9$}
 \psfrag{M9c}{\Large$\mathcal{M}^c_9$}
\psfrag{M9l}{\Large$\mathcal{M}^l_9$}
\psfrag{M9r}{\Large$\mathcal{M}^r_9$}

\centerline{\includegraphics[width=160mm,height=170mm]{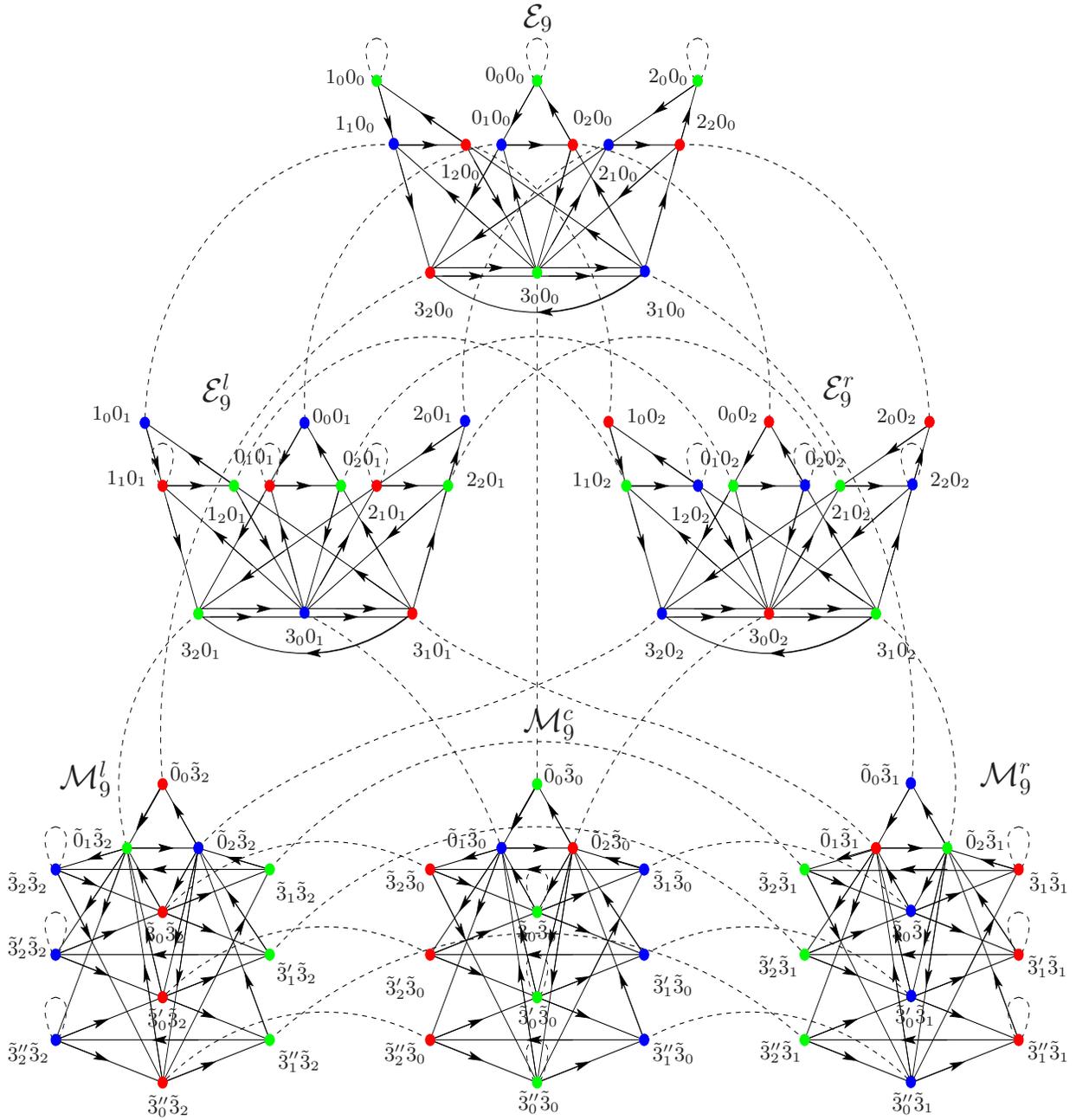}} 
\caption{The Ocneanu graph $Oc(\mathcal{E}_9) = Oc(\mathcal{M}_9)$. The two left chiral generators are $0_1 \otimes 0_0$ and $0_2 \otimes 
0_0$, the two right chiral generators are $0_0 \otimes 0_1$ and $0_0 \otimes 0_2$. The tensor product $a \otimes b$ is denoted with the shorthand notation $ab$.}
\label{GraphOcE9}
\end{figure}

\paragraph{The Ocneanu algebra of quantum symmetries and a realization}
The matrix $V_{(1,0),(0,0)}$ is the adjacency matrix of the left chiral part of the Ocneanu graph. The graph is composed of six subgraphs, three copies of the $\mathcal{E}_9$ graph and three copies of the $\mathcal{M}_9$ graph, as showed on figure \ref{GraphOcE9}. We label the vertices as follows: $x=a\otimes 0_i$ with $a, 0_i \in \mathcal{E}_9$ and $i=0,\,1,\,2$ for vertices of $\mathcal{E}_9$-type subgraphs and  $x=\tilde{a} \otimes \tilde{3}_i$ with $\tilde{a}, \tilde{3}_i \in\mathcal{M}_9$ and $i=0,\,1,\,2$ for vertices of $\mathcal{M}_9$-type subgraphs.
The matrix $V_{(1,0),(0,0)}$ corresponds to the multiplication by the left chiral generator $0_1 \otimes 0_0$. The matrix $V_{(0,0),(1,0)}$ is the adjacency matrix of the right chiral part of the Ocneanu graph $Oc(\mathcal{E}_9)$, and corresponds to the multiplication by the right chiral generator $0_0 \otimes 0_1$. The dashed lines in the graph corresponds to the chiral operator $C$. We have 
$V_{(0,0),(1,0)} = C V_{(1,0),(0,0)} C^{-1}$. The multiplication by $0_0 \otimes 0_1$ is obtained as follows. We start with $x$, apply $C$, multiply the result by $0_1 \otimes 0_0$, and apply $C^{-1}=C$. 
From matrices $V_{(1,0),(0,0)}$ and $V_{(0,0),(1,0)}$ all others $V_{\lambda\mu}$ (hence also the double toric matrices $W_{xy}$) are calculated straightforwardly using equations (\ref{Vfusion1}--\ref{Vfusion3}).

From the multiplication by chiral left and right generators $0_1 \otimes 0_0$ and $0_0 \otimes 0_1$ (and their conjugate) we reconstruct the multiplication table of $Oc(\mathcal{E}_9)$. As for the graph matrices of $\mathcal{E}_9$, the calculation is not straightforward, but imposing non-negative integer coefficients leads to a unique solution. The result is encoded in the 72 quantum symmetry matrices $O_x$ of dimension $72 \times 72$.  

\paragraph{Realization of the quantum symmetry algebra} 
In order to have a compact (readable) description of these matrices and the multiplicative structure of the algebra of quantum symmetries, we propose the following realization of this algebra: 
\begin{equation}
Oc \, = \, ``\mathcal{E}_9 \otimes_{\mathbb{Z}_3} \mathcal{E}_9 \textrm{''} \, \stackrel{\cdot}{=} \, 
(\mathcal{E}_9 \otimes_{\rho} \mathcal{E}_9) \oplus (\mathcal{M}_9 \otimes_{\rho} \mathcal{M}_9) \;,
\end{equation}
where the notation $\otimes_{\rho}$ means that the tensor product is quotiented using the $\mathbb{Z}_3$ symmetry of graphs $\mathcal{E}_9$ and $\mathcal{M}_9$ in the following way. A basis of the quantum symmetry algebra is given by elements $\{a \otimes 0_i \, , \, \tilde{a} \otimes \tilde{3}_i \}$ for $i=0,1,2$. The other elements of $\mathcal{E}_9 \otimes \mathcal{E}_9$ and $\mathcal{M}_9 \otimes \mathcal{M}_9$ are identified with basis elements $\{a \otimes 0_i \, , \, \tilde{a} \otimes \tilde{3}_i \}$ using the $\mathbb{Z}_3$ symmetry  operators $\rho_3$ and $\rho_3'$ of graphs 
$\mathcal{E}_9$ and $\mathcal{M}_9$ and the induction-restruction rules between the two graph algebras, as follows:
\begin{eqnarray}
\label{proj1} \bullet \qquad a \otimes 1_i &=& a\otimes 1_0 \cdot 0_i \,=\, 1_0\cdot a\otimes 0_i \,=\, \rho_3(a) \otimes 0_i \\
\bullet \qquad a \otimes 2_i &=& a\otimes 2_0 \cdot 0_i \,=\, 2_0\cdot a\otimes 0_i \,=\, (\rho_3)^2(a) \otimes 0_i \\
\bullet \qquad a \otimes 3_i &=& \sum_{\tilde{a}} (E_{\tilde{0}_0})_{a\tilde{a}}\, \tilde{a} \otimes \tilde{3}_i \\
\bullet \qquad \tilde{a} \otimes \tilde{3}_i' &=& \tilde{a} \otimes 1_0 \cdot \tilde{3}_i = 1_0 \cdot \tilde{a} \otimes \tilde{3}_i = \rho_3'(\tilde{a}) \otimes \tilde{3}_i \\ 
\bullet \qquad \tilde{a} \otimes \tilde{3}_i'' &=& \tilde{a} \otimes 2_0 \cdot \tilde{3}_i = 2_0 \cdot \tilde{a} \otimes \tilde{3}_i = (\rho_3')^2(\tilde{a}) \otimes \tilde{3}_i \\
\label{proj6} \bullet \qquad \tilde{a} \otimes \tilde{0}_i &=& \sum_{a}(E^{tr}_{\tilde{0}_0})_{\tilde{a},a}\; a \otimes 0_i
\end{eqnarray}
Here the matrix $E_{\tilde{0}_0}$ encodes the branching rules $\mathcal{E}_9 \hookrightarrow \mathcal{M}_9$ (obtained from matrices $P^{\ell}$ implementing the $\mathcal{E}_9$ (left) action on $\mathcal{M}_9$ as follows: $(E_{\tilde{b}})_{a\tilde{c}} = (P_a^{\ell})_{\tilde{b}\tilde{c}}$). Explicitely, we have:
{\scriptsize \begin{equation}
E_{\tilde{0}_0} = \left(
\begin{array}{cccccccccccc}
1& .& .& .& .& .& .& .& .& .& .& . \\
1& .& .& .& .& .& .& .& .& .& .& . \\ 
1& .& .& .& .& .& .& .& .& .& .& . \\ 
.& 1& 1& 1& .& .& .& .& .& .& .& . \\ 
.& .& .& .& 1& .& .& .& .& .& .& . \\ 
.& .& .& .& 1& .& .& .& .& .& .& . \\
.& .& .& .& 1& .& .& .& .& .& .& . \\ 
.& .& .& .& .& 1& 1& 1& .& .& .& . \\ 
.& .& .& .& .& .& .& .& 1& .& .& . \\ 
.& .& .& .& .& .& .& .& 1& .& .& . \\ 
.& .& .& .& .& .& .& .& 1& .& .& . \\ 
.& .& .& .& .& .& .& .& .& 1& 1& 1
\end{array}
\right)
\qquad \qquad
\begin{array}{rcl}
0_0 &\hookrightarrow& \tilde{0}_0 \\
1_0 &\hookrightarrow& \tilde{0}_0 \\
2_0 &\hookrightarrow& \tilde{0}_0 \\
3_0 &\hookrightarrow& \tilde{3}_0 + \tilde{3}_0' + \tilde{3}_0''  \\
0_1 &\hookrightarrow& \tilde{0}_1 \\
1_1 &\hookrightarrow& \tilde{0}_1 \\
2_1 &\hookrightarrow& \tilde{0}_1 \\
3_1 &\hookrightarrow& \tilde{3}_1 + \tilde{3}_1' + \tilde{3}_1'' \\
0_2 &\hookrightarrow& \tilde{0}_2 \\
1_2 &\hookrightarrow& \tilde{0}_2 \\
2_2 &\hookrightarrow& \tilde{0}_2 \\ 
3_2 &\hookrightarrow& \tilde{3}_2 + \tilde{3}_2' + \tilde{3}_2'' 
\end{array}
\end{equation}}
\normalsize
The multiplication of the basis generators $\{a \otimes 0_i \, , \, \tilde{a} \otimes \tilde{3}_i \}$ is then naturally  defined using the multiplication rules (\ref{multh}) and the projections (\ref{proj1}--\ref{proj6}). We introduce the matrices 
$R^{r}$ defined from the right action of $\mathcal{E}_9$ on $\mathcal{M}_9$: 
$\tilde{b} \, a = \sum_{\tilde{c}} (P^{r}_a)_{\tilde{b}\tilde{c}} \, \tilde{c} = \sum_{\tilde{c}} (R^{r}_{\tilde{b}})_{a\tilde{c}} \, \tilde{c}$. It can be seen that the algebra $Oc(\mathcal{E}_9)$ is non commutative and isomorphic with the direct sum of 9 copies of $2\times 2$ matrices and 36 copies of the complex numbers. With our parametrisation, the quantum symmetry matrices read:
{\scriptsize
\begin{equation}
\begin{array}{rcc}
O_{a\otimes0_0} 
&=&
\left(
\begin{array}{cccccc}
G_a & . & . & . & . & . \\
  . &G_a& . & . & . & . \\
  . & . &G_a& . & . & . \\
  . & . & . &P_a^{\ell}& . & . \\
  . & . & . & . &P_a^{\ell}& . \\
  . & . & . & . & . &P_a^{\ell}\\
\end{array}
\right) \\
{} & {} \\
O_{a\otimes 0_1}
&=&
\left(
\begin{array}{cccccc}
.   & G_a            & .             & .             & .                  & .\\
.   & .              & G_a           & .             & .                  & G_a\,E_0 \\
G_a & .              & .             & G_a\,E_0  & .                  & . \\
.   & P_a^{\ell}\,E^{tr}_0 & .             & .             & P_a^{\ell}\,(\munite+ P_{1_0}^{\ell}) & . \\
.   & .              & P_a^{\ell}\,E^{tr}_0& .             & .                  & P_a^{\ell} \\
.   & .              & .             & P_a^{\ell}\,(\munite + P_{2_0}^{\ell})& .                  &.\\
\end{array}
\right) \\
{} & {} \\
O_{a\otimes 0_2}
&=&
\left(
\begin{array}{cccccc}
.       & .              & G_a           & .             & .                      & .      \\
G_a     & .              & .             & G_a\,E_0        & .                      &.       \\
.       & G_a            & .             & .             & G_a\,E_0                 & .      \\
.       & .              & P_a^{\ell}\,E^{tr}_0      & .             & .                      & P_a^{\ell} \,(\munite+ 
P_{1_0}^{\ell}) \\
.       & .              & .             &  P_a^{\ell}\,(\munite + P_{2_0}^{\ell})&  .  & .       \\
.       & P_a^{\ell}\,E^{tr}_0       & .             & .             & P_a^{\ell}                    & .       \\
\end{array}
\right) \\
{} & {} \\
O_{\tilde{a}\otimes\tilde{3}_0}
&=&
\left(
\begin{array}{cccccc}
  .  &  .               &  .  &R^r_{\tilde{a}}&  .  &  .  \\
  .  &R^r_{\tilde{a}}\,E_0^{tr} &  .  & .           & R^r_{\tilde{a}}(\munite+ P^{\ell}_{1_0}) & . \\
  .  &  .  & R^r_{\tilde{a}}E_0^{tr} & . & . &  R^r_{\tilde{a}}(\munite+P^{\ell}_{1_0}) \\
 H_{\overline{a}} & . & . & 2(H_{\tilde{a}}E_0) & . & . \\
 .   & H_{\tilde{a}}(\munite+G_{1_0})&. &. & H_{\tilde{a}}E_0 & . \\
. & . & H_{\tilde{a}}(\munite+ G_{1_0}) & . & . & H_{\tilde{a}}E_0
\end{array}\right) \\
{} & {} \\
O_{\tilde{a}\otimes\tilde{3}_1}
&=&\left(
\begin{array}{cccccc}
.  &  .  &  .  &  .  & R^r_{\tilde{a}} & .   \\
.  &  .  & R^r_{\tilde{a}} \,E_0^{tr}& . & . & R^r_{\tilde{a}} \\
. & . & . & R^r_{\tilde{a}}(\munite + P^{\ell}_{2_0}) & . & . \\
. & H_{\tilde{a}}(\munite + G_{2_0}) &. &. &H_{\tilde{a}}\,E_0 & . \\
. & . & H_{\tilde{a}} & . & . & H_{\tilde{a}}\,E_0 \\
H_{\tilde{a}}& . & . & H_{\tilde{a}}\,E_0 & . & . \\
\end{array}
\right) \\
{} & {} \\
O_{\tilde{a}\otimes\tilde{3}_2}
&=&
\left(
\begin{array}{cccccc}
. & . & . & . & . & R^r_{\tilde{a}} \\
. & . & . & R^r_{\tilde{a}}(\munite + P^{\ell}_{2_0}) &. &. \\
. & R^r_{\tilde{a}} \, E^{tr}_0 & . & . &  R^r_{\tilde{a}} & . \\
. & . & H_{\tilde{a}}(\munite+G_{2_0}) & . & . & H_{\tilde{a}}\,E_0 \\
H_{\tilde{a}} & . & . & H_{\tilde{a}}\,E_0 & . & . \\
. & H_{\tilde{a}} & . & . & H_{\tilde{a}}\,E_0 & .
\end{array}
\right)
\end{array}
\end{equation}}

Triality $t$ is well defined on this algebra: $t(a_i \otimes 0_j) = t(\tilde{a}_i \otimes \tilde{3}_j ) = i+j \,(\textrm{mod} 3)$. The left chiral subalgebra (by definition the algebra generated by the left chiral generator $0_1 \otimes 0_0$) is $L = \{ a \otimes 0_0 \}$. The right chiral subalgebra (generated by $0_0 \otimes 0_1$) is $R = \{ 0_0 \otimes a \}$. With the projections (\ref{proj1}-\ref{proj6}), $R$ correspondonds to the set of elements $\{ 0_0 \otimes 0_0, 1_0 \otimes 0_0, 2_0 \otimes 0_0, \tilde{0}_0 \otimes \tilde{3}_0, 0_0 \otimes 0_1, 1_0 \otimes 0_1, 2_0 \otimes 0_1, \tilde{0}_0 \otimes \tilde{3}_1, 0_0 \otimes 0_2, 1_0 \otimes 0_2, 2_0 \otimes 0_2, \tilde{0}_0 \otimes \tilde{3}_2 \}$. The ambichiral subalgebra (by definition the intersection of $L$ and $R$) is $A = \{ 0_0 \otimes 0_0, 1_0 \otimes 0_0, 2_0 \otimes 0_0 \}$. The chiral operation $C$ on the basis elements is defined by $C(u \otimes v) = (v \otimes u)$, for $u,v \in H = \mathcal{E}_9 \oplus \mathcal{M}_9$ (and using the projections (\ref{proj1}-\ref{proj6})). The self-dual elements obey $C(u)=u$, they are the ones in figure \ref{GraphOcE9} which are connected to themselves by the dashed line. $A$-elements are, in particular, self-dual.

\paragraph{One modular invariant and two graphs}
Starting from the modular invariant (\ref{modinve9}), we obtain the set of toric matrices $W_{x0}$, double fusion matrices $V_{\lambda\mu}$ and quantum symmetry matrices $O_x$, together with the corresponding Ocneanu graph. By an analysis of the latter, it clearly appears that there are two graphs that are modules under the quantum symmetry algebra, the $\mathcal{E}_9$ and $\mathcal{M}_9$ graphs. 
Using the realization of the quantum symmetry algebra described above, the module structure for $\mathcal{E}_9$ is defined by:
\begin{equation}
Oc \times \mathcal{E}_9 \rightarrow \mathcal{E}_9 \qquad \qquad \qquad
\left\lbrace \begin{array}{rcl} 
(a \otimes 0_i) \, \cdot \, b & \doteq & a \, \cdot \, b \, \cdot \, 0_i  = a \, \cdot \, 0_i \, \cdot \, b  \\ 
(\tilde{a} \otimes \tilde{3}_0) \, \cdot \, b  & \doteq & \tilde{a} \, \cdot \, b \, \cdot \, \tilde{3}_0 \\
(\tilde{a} \otimes \tilde{3}_{1,2}) \, \cdot \, b  & \doteq & \tilde{a} \, \cdot \, \rho(b) \, \cdot \, \tilde{3}_{1,2} = \tilde{a} \, \cdot \, b \, \cdot \, \rho'(\tilde{3}_{1,2}) \\
\end{array}
\right.
\end{equation} 
and the corresponding dual annular matrices are:
\begin{equation}
S_{x=a\otimes 0_i}^{\mathcal{E}} = G_{0_i} \, G_a \;, \qquad  \qquad
S_{x=\tilde{a} \otimes \tilde{3}_0}^{\mathcal{E}} = L_{\tilde{3}_0} \, H_{\tilde{a}} \;, \qquad \qquad
S_{x=\tilde{a} \otimes \tilde{3}_{1,2}}^{\mathcal{E}} = L_{\rho'(\tilde{3}_{1,2})} \, H_{\tilde{a}} \;,
\end{equation}
where the $L_{\tilde{b}}$ matrices are defined by $a \cdot \tilde{b} = \sum_{\tilde{c}} (L_{\tilde{b}})_{a\,\tilde{c}} \, \tilde{c}$.  
The module structure for $\mathcal{M}_9$ is defined by:
\begin{equation}
Oc \times \mathcal{M}_9 \rightarrow \mathcal{M}_9 \qquad 
\left\lbrace \begin{array}{rcl} 
(a \otimes 0_i) \, \cdot \, \tilde{b} & \doteq & a \, \cdot \, \tilde{b} \, \cdot \, 0_i = a \, \cdot \, 0_i \, \cdot \, \tilde{b} \\ 
(\tilde{a} \otimes \tilde{3}_i) \, \cdot \, \tilde{b}  & \doteq & \tilde{a} \, \cdot \, \tilde{b} \, \cdot \, \tilde{3}_i
\end{array}
\right.
\end{equation} 
and the corresponding dual annular matrices are:
\begin{equation}
S_{x = a \otimes 0_i}^{\mathcal{M}} = P_{0_i}^{\ell} \, P_{a}^{\ell} \;, \qquad \qquad \qquad 
S_{x = \tilde{a} \otimes \tilde{3}_i}^{\mathcal{M}} = H_{\tilde{a}} \, L_{\tilde{3}_i} \;.
\end{equation}

We have therefore two quantum groupo\"ids associated with the initial modular invariant, constructed from the graphs $\mathcal{E}_9$ and $\mathcal{M}_9$. Setting $d_{\lambda}^{\mathcal{E}} = \sum_{a,b} (F_{\lambda}^{\mathcal{E}})_{ab}$, $d_{x}^{\mathcal{E}} = \sum_{a,b} (S_{x}^{\mathcal{E}})_{ab}$,
$d_{\lambda}^{\mathcal{M}} = \sum_{a,b} (F_{\lambda}^{\mathcal{M}})_{ab}$, $d_{x}^{\mathcal{M}} = \sum_{a,b} (S_{x}^{\mathcal{M}})_{ab}$, we check 
the dimensional rules:
\begin{eqnarray}
\dim(\mathcal{B}(\mathcal{E}_9) &=& \sum_{\lambda} (d_{\lambda}^{\mathcal{E}})^2 = \sum_x (d_x^{\mathcal{E}})^2 = 518\,976 \;. \\
\dim(\mathcal{B}(\mathcal{M}_9) &=& \sum_{\lambda} (d_{\lambda}^{\mathcal{M}})^2 = \sum_x (d_x^{\mathcal{M}})^2 = 754\,272 \;.
\end{eqnarray}

\begin{figure}
\psfrag{1}{\footnotesize$\widehat0_0$}
  \psfrag{2}{\footnotesize$\widehat3_0$}
  \psfrag{3}{\footnotesize$\widehat3'_0$}
  \psfrag{4}{\footnotesize$\widehat3''_0$}
  \psfrag{5}{\footnotesize$\widehat0_1$}
  \psfrag{6}{\footnotesize$\widehat3_1$}
  \psfrag{7}{\footnotesize$\widehat3'_1$}
  \psfrag{8}{\footnotesize$\widehat3''_1$}
  \psfrag{9}{\footnotesize$\widehat0_2$}
  \psfrag{10}{\footnotesize$\widehat3_2$}
  \psfrag{11}{\footnotesize$\widehat3'_2$}
  \psfrag{12}{\footnotesize$\widehat3''_2$}
\centerline{\includegraphics[width=60mm,height=60mm]{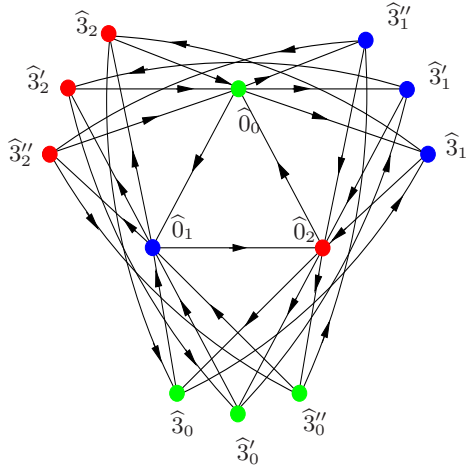}} 
\caption{The rejected Di Francesco-Zuber graph.}
\label{gr-zub}
\end{figure}

\paragraph{The rejected diagram}
In the first list of $SU(3)$-type graphs presented by Di Francesco and Zuber in \cite{DiFZuber}, there were three graphs associated with the exceptional modular invariant (\ref{modinve9}): the graphs $\mathcal{E}_9$, $\mathcal{M}_9$ and the one displayed on figure \ref{gr-zub}, denoted $\mathcal{Z}_9$. This graph was later rejected by Ocneanu in \cite{Oc-Bariloche} because some required cohomological property (written in terms of values for triangular cells) was not fullfilled. In other words, this graph gives rise to a module over the ring of $\mathcal{A}_9$, with the right properties, but the underlying category does not exist. 

In this paper, the higher Coxeter graphs are obtained as subgraphs or module graphs of their Ocneanu graph, which encodes the quantum symmetry algebra $Oc(G)$ previously determined. For Type I partition functions (block diagonal with respect to the characters of the extended chiral algebra) the associated graphs have self-fusion, they appear directly as subgraphs of their Ocneanu graph (this is the case, for instance, for the $\mathcal{E}_5$ and $\mathcal{E}_9$ graphs presented here). For Type II partition functions, the associated graphs are called ``module'' graphs. They define a module over $Oc$, but they are most easily determined as a module over a self-fusion subgraph of the Ocneanu graph, called its parent graph. 
For all $su(3)$ cases studied, module graphs can be obtained from orbifold or conjugation methods from their parent graph. This is indeed the case for the conjugate $\mathcal{A}$ series and the orbifold and conjugate orbifold series $\mathcal{D}$ and $\mathcal{D}^*$. This is also the case for the $\mathcal{E}_5/3$ and $\mathcal{M}_9 = \mathcal{E}_9/3$ graphs. There is also the exceptional twist, but in this case the graph appears directly as a subgraph of its Ocneanu graph (see \cite{Dah-tesis}). In the particular case of the graph displayed on figure \ref{gr-zub}, the graph can not be obtained from $\mathcal{E}_9$ by orbifold or conjugation methods, and this fact may indicate a hint that such graph should be rejected. 

Nevertheless, let us present some properties of this graph. The vector space of $\mathcal{Z}_9$ is a module over the left-right action of $\mathcal{A}_9$, encoded by the annular matrices $\mathcal{F}_{\lambda}^{\mathcal{Z}}$ computed as usual from the recursion relation (\ref{su3recrel}) with $F^{\mathcal{Z}}_{(0,0)}=\munite$, $F^{\mathcal{Z}}_{(1,0)} = Ad(\mathcal{Z}_9)$. The vector space of $\mathcal{Z}_9$ is also a module over the left action of the $\mathcal{E}_9$ graph, encoded by the set of matrices $D_{a}$
\begin{equation}
\mathcal{E}_9 \times \mathcal{Z}_9 \rightarrow \mathcal{Z}_9 : \quad a \cdot \hat{b} = 
 \sum_{\hat{c}} (D_{a})_{\hat{b}\hat{c}}\;\hat{c} \qquad \qquad a \in \mathcal{E}_9\;, \quad 
\hat{b},\hat{c} \in \mathcal{Z}_9 \;.
\end{equation}
We compute the set of matrices $D_{a}$ using the multiplicative structure of $\mathcal{E}_9$ as previously.   
In the ordered basis $(\hat{0}_0,\hat{3}_0,\hat{3}_0',\hat{3}_0'';  
\hat{0}_1,\hat{3}_1,\hat{3}_1',\hat{3}_1''; \hat{0}_2,\hat{3}_2,\hat{3}_2',\hat{3}_2'')$, the matrices $D_{1_0}$ and $D_{2_0}$ are given by the same matricial expression as in (\ref{pEl}). 
The vector space of $\mathcal{Z}_9$ is also a $Oc$-module. Using the realization of the quantum symmetry algebra, the action is defined by:
\begin{equation}
Oc \times \mathcal{Z}_9 \rightarrow \mathcal{Z}_9 \qquad \quad 
\left\lbrace \begin{array}{rcl} 
(a \otimes 0_i) \, \cdot \, \hat{b} &\doteq& a \, \cdot \, 0_i \, \cdot \, \hat{b} \\ 
(\tilde{a} \otimes \tilde{3}_0) \, \cdot \, \hat{b}  &\doteq& (\tilde{a} \, \cdot \, \tilde{3}_0 )\, \cdot \, t(\hat{b}) \\
(\tilde{a} \otimes \tilde{3}_{1,2}) \, \cdot \, \hat{b}  &\doteq& (\tilde{a} \, \cdot \, \rho'(\tilde{3}_{1,2}) )\, \cdot \, t(\hat{b})
\end{array}
\right.
\end{equation}
where the operator $t$ is defined on the vertices of $\mathcal{Z}_9$ by $t(\hat{0}_i)=\hat{0}_i, t(\hat{3}_i)=\hat{3}_i, t(\hat{3}_i')=\hat{3}_i'', t(\hat{3}_i'')=\hat{3}_i'$. We also define the matrices $D_{a}^t$ by the relations $(D_a^t)_{\hat{b}\hat{c}} = (D_a)_{t(\hat{b})\hat{c}} \,$. The quantum symmetry matrices for $\mathcal{Z}_9$ are:
\begin{equation} 
S_{x=a\otimes 0_i}^{\mathcal{Z}} = D_{0_i} \, D_a \;, \qquad 
S_{x=\tilde{a} \otimes \tilde{3}_0}^{\mathcal{Z}} =  \sum_c (H_{\tilde{a}})_{\tilde{3}_0 \,c} \, D_c^{t} \;, \qquad 
S_{x=\tilde{a} \otimes \tilde{3}_{1,2}}^{\mathcal{Z}} =  \sum_c (H_{\tilde{a}})_{\rho'(\tilde{3}_{1,2}) \,c} \, D_c^{t} \;.
\end{equation} 
We can also check the dimensional rules:
$$
 \sum_{\lambda} (d_{\lambda}^{\mathcal{Z}})^2 = \sum_x (d_x^{\mathcal{Z}})^2 = 754 \, 272 \;.
$$
Therefore, the graph $\mathcal{Z}_9$ satisfy all module properties and dimensional rules. 
Even if it does not appear directly as a byproduct of the calculations presented in this paper (giving a hint for its rejection), its formal rejection only seems possible with additional data of cohomological nature (cells), by CFT arguments or in the subfactor approach.

\paragraph{Final Comment}
The Ocneanu graphs displayed in this paper ($Oc(\mathcal{E}_5)$, $Oc(\mathcal{E}_9)$) have been first obtained by Ocneanu himself. For instance those associated with members of the $su(3)$ family were displayed on posters during the Bariloche conference (2000) but the full list never appeared in print. Several techniques \cite{Coq_Gil-Tmod,Gil-tesis} allow one to recover some of them from the knowledge of the Di Francesco - Zuber diagrams. The present paper actually emerged from our wish to obtain the Ocneanu graphs $Oc(G)$ (and the graphs $G$ themselves, of course) from the only data provided by the modular invariant. 

\paragraph{Acknowledments}
We thank the referee for his constructive remarks and for bringing to our attention the reference \cite{Xu}.
We also wish to thank R. Coquereaux for his suggestions, guidance, and help. G. Schieber was supported by a fellowship of Agence Universitaire de la Francophonie (AUF) and of FAPERJ, and thanks IMPA for its hospitality during the final corrections of the paper.


{}

\end{document}